\documentclass[12pt, draftclsnofoot, onecolumn]{IEEEtran}
\usepackage{amsmath}
\usepackage{amssymb}
\usepackage{amsthm}
\usepackage{cite}
\usepackage{color}
\usepackage{xcolor}
\usepackage{caption}
\usepackage{epsfig,latexsym}
\usepackage{float}
\usepackage{fancyhdr}
\usepackage{graphicx}
\usepackage{indentfirst}
\usepackage{mdwmath}
\usepackage{mdwtab}
\usepackage{subfigure}
\usepackage{setspace}
\usepackage{times}
\usepackage{url}

\newtheorem{remark}{Remark}
\newtheorem{theorem}{Theorem}

\newtheorem{lemma}{Lemma}

\newtheorem{corollary}{Corollary}

\newtheorem{proposition}{Proposition}

\begin{document}

\title{Multiple Antenna Aided NOMA in UAV Networks: A Stochastic Geometry Approach}
\author{Tianwei Hou,~\IEEEmembership{Student Member,~IEEE,}
        Yuanwei Liu,~\IEEEmembership{Member,~IEEE,}
        Zhengyu Song,
        Xin Sun,
        and Yue Chen,~\IEEEmembership{Senior Member,~IEEE,}

\thanks{This work is supported by the Fundamental Research Funds for the Central Universities under Grant 2016RC055. This paper was presented at the IEEE Global Communication Conference, Abu Dhabi, United Arab Emirates, Dec. 2018~\cite{Hou_UAV_Glob}.}
\thanks{T. Hou, Z. Song and X. Sun are with the School of Electronic and Information Engineering, Beijing Jiaotong University, Beijing 100044, China (email: 16111019@bjtu.edu.cn, songzy@bjtu.edu.cn, xsun@bjtu.edu.cn).}
\thanks{Y. Liu and Yue Chen are with School of Electronic Engineering and Computer Science, Queen Mary University of London, London E1 4NS, U.K. (e-mail: yuanwei.liu@qmul.ac.uk, yue.chen@qmul.ac.uk).}
}

\maketitle

\begin{abstract}
This article investigates the multiple-input multiple-output (MIMO) non-orthogonal multiple access (NOMA) assisted unmanned aerial vehicles (UAVs) networks. By utilizing a stochastic geometry model, a new 3-Dimension UAV framework for providing wireless service to randomly roaming NOMA users has been proposed. In an effort to evaluate the performance of the proposed framework, we derive analytical expressions for the outage probability and the ergodic rate of MIMO-NOMA enhanced UAV networks. We examine tractable upper bounds for the whole proposed framework, with deriving asymptotic results for scenarios that transmit power of interference sources being proportional or being fixed to the UAV. For obtaining more insights for the proposed framework, we investigate the diversity order and high signal-to-noise (SNR) slope of MIMO-NOMA assisted UAV networks. Our results confirm that: i) The outage probability of NOMA enhanced UAV networks is affected to a large extent by the targeted transmission rates and power allocation factors of NOMA users; and ii) For the case that the interference power is proportional to the UAV power, there are error floors for the outage probabilities.
\end{abstract}


\section{Introduction}

\IEEEPARstart{I}{n} the fifth generation (5G) communication systems and internet of things (IoT) networks, massive connectivity is required to support large number of devices in various scenarios with limited spectrum resources~\cite{wireless_sparse,wideband_Qin}. Unmanned aerial vehicles (UAVs) communication is an effective approach to provide connectivity during temporary events and after disasters. Some initial research contributions in the field of UAVs communication have been made by researchers in~\cite{Saad_D2D_UAV,UAV_zeng,Xiao_mm_wave_UAV,smallUAV_frew,Saad_LoS_coverage}. UAVs can provide connectivity to multiple users as wireless relays for improving wireless coverage. On the other hand, ground users can have access to UAVs, which can act as aerial base stations, for reliable downlink and uplink communications~\cite{Saad_D2D_UAV,UAV_zeng,Xiao_mm_wave_UAV}.
Additionally, UAV networks are capable of enhancing spectrum efficiency by having line-of-sight (LoS) connections towards the users, which provide higher received power for the users~\cite{smallUAV_frew,Saad_LoS_coverage}.% In order to further exploit the spectrum efficiency of UAV networks, non-orthogonal multiple access (NOMA) is a significant solution.

Non-orthogonal multiple access (NOMA) has been considered as a promising technique in 5G mobile networks because of its superior spectrum efficiency~\cite{NOMA_5G_beyond_Liu,NOMA_mag_DingLiu}. The key idea is that multiple users are served within the same frequency, time and code block~\cite{modulation_Cai,heterNOMA_Qin}, which relies on the employment of superposition coding (SC) and successive interference cancellation (SIC) at the transmitter and receiver, respectively~\cite{Saito2013}. Moreover, SIC technique at receivers allows that users who have better channel conditions to remove the intra-channel interference. Ding {\em et al.}~\cite{PairingDING2016} estimated the performance of NOMA with fixed power allocation (F-NOMA) and cognitive radio inspired NOMA, which demonstrated that only the user with higher channel gain influences the system performance. Timotheou and Krikidis optimized the user-power allocation problem to improve user fairness of a NOMA system \cite{Timo_userfairness}. To further exploit NOMA networks, the outage performance and capacity in the downlink and uplink transmission scenario with dynamic power allocation factors was estimated by Yang {\em et al.}~\cite{General_power_DING}, which can guarantee the quality of service in dynamic power allocation NOMA (D-NOMA).

\subsection{Motivations and Related Works}

1) \emph{Studies on MIMO-NOMA systems:}  Current multiple-input multiple-output (MIMO) NOMA applications can be primarily classified into a pair of categories, namely, beamformer based structure and cluster based structure. In beamformer based structure, one centralized beamformer serves only one user in a beam~\cite{Liu_MIMO}. Choi proposed a coordinated multi-point transmission scheme in~\cite{beamformer_based_Choi}, where two base stations (BS) serve paired NOMA users simultaneously by beamformer based structure. The multiple antenna scenario of beamformer based structure was proposed by Shin {\em et al.} in~\cite{MIMO_beamformer_based}, where a joint centralized and coordinated beamforming design was developed for suppressing the inter-cell interference. Some related works on cluster based MIMO-NOMA have been investigated in~\cite{P_identity_journal_DING,MIMO_journal_DING_alignment,Liu_Coop_NOMA_SWIPT,Liu_Coop_NOMA_SWIPT,Liu_physical_scurity_NOMA}. More specifically, Ding {\em et al.}~\cite{P_identity_journal_DING} proposed a MIMO-NOMA model with transmit power control and detection scheme. By adopting this design, the MIMO-NOMA model can be separated to multiple independent SISO-NOMA arrangements. Then, outage probabilities and the sum rate gap of multiple SISO-NOMA arrangements were estimated, which the model assumes that global channel state information (CSI) is unknown at the BS. The sum rate and multi-user capacity were investigated by Zeng {\em et al.}~\cite{Capacity_Zeng}, which show that multi-user MIMO-NOMA is not a preferable solution of NOMA due to high computational complexity. In order to establish a more general framework of MIMO-NOMA, and circumvent the restrictive assumption of receiver antennas, the signal alignment technique was proposed for both downlink and uplink transmission scenarios in~\cite{MIMO_journal_DING_alignment}.

2) \emph{Studies on NOMA in stochastic geometry systems:} Stochastic geometry is an effective mathematical tool for capturing the topological randomness of networks. As such, stochastic geometry tools were invoked to model the impact of the locations for NOMA users~\cite{Randomly_ding}. Some research contributions with utilizing stochastic geometry approaches have been studied in~\cite{Liu_Coop_NOMA_SWIPT,Liu_physical_scurity_NOMA}. More particularly, Liu {\em et al.}~\cite{Liu_Coop_NOMA_SWIPT} proposed an innovative model of cooperative NOMA with simultaneous wireless information and power transfer (SWIPT). In this model, the wireless power transfer technique was employed at users, where near users acted as energy harvesting relays for supporting far users. Ding {\em et al.}~\cite{Randomly_ding} evaluated the performance of NOMA with randomly deployed users. The analytical results show that it is more preferable to group users whose channel gains are more distinctive to improve the diversity order in NOMA system. With the goal of enhancing the physical layer security of NOMA networks, Liu \emph{et al.}~\cite{Liu_physical_scurity_NOMA} proposed a NOMA assisted physical layer security framework in large-scale networks. The secrecy performance of both single antenna and multiple antenna aided BS scenarios have been investigated.

3) \emph{Studies on UAV:} In UAV aided wireless networks, the probability that each device has a LoS link is dependent on the environment, location of the device, carrier frequency and the elevation angle~\cite{Saad_D2D_UAV,Saad_LoS_coverage,Hourani_UAV_Altitude}.
%In general cases, the ground users are exposed to UAVs, so that the LoS links between UAVs and users are expected for UAV-aided communications. Sporadically, ground users are not expected in the open areas, they could also be blocked by obstacles such as buildings or terrain. The UAV-ground channels may also affected by a number of multi-path components due to the reflection, scattering and diffraction by obstacles. Therefore, one must consider the randomness associated with the LoS and non-line-of-sight (NLoS) links while designing the UAV-based communication system.
Sharma and Kim estimated the outage performance of single antenna assisted NOMA in UAV networks~\cite{Kim_NOMA_UAV}, where UAV-ground channels are characterized by LoS transmission. Recently, the general form of both LoS and NLoS transmission scenarios, namely, Nakagami-\emph{m} fading channels, have been proposed in the literature. Hou {\em et al.}~\cite{Nakagami_Hou} estimated the outage performance of F-NOMA downlink transmission scenario in both LoS and NLoS scenarios. %Additionally, a MIMO-NOMA aided UAV framework, where stochastic geometry tools were invoked to estimate the outage performance, was proposed by Hou {\em et al.}~\cite{Hou_UAV_Glob}.

While the aforementioned research contributions have laid a solid foundation with providing a good understanding of single input single output (SISO) NOMA terrestrial networks, how MIMO-NOMA technique is capable of assisting UAV networks is still unknown. To the best of our knowledge, there has been no existing work intelligently investigating the effect of the network performance of MIMO-NOMA assisted UAV networks, which motivates us to develop this treatise.
\vspace{-0.1in}
\subsection{Contributions}
The novel structure design in this work--by introducing the MIMO-NOMA assisted UAV networks--can be a new highly rewarding candidate, which will contribute the following key advantages:
\begin{itemize}
  \item We propose a general MIMO-NOMA aided UAV framework with interference, where stochastic geometry approaches are invoked to model the locations of users and interference sources. Utilizing this framework, LoS and NLoS links are considered to illustrate the general case of NOMA assisted UAV networks.
  \item We derive closed-form expressions for outage probability of paired NOMA users in the proposed framework. We provide tractable analytical upper bounds for both LoS and NLoS scenarios. Diversity orders are obtained for the paired NOMA users based on the developed outage probability. The obtained results confirm that for the case that interference power is proportional to the transmit power, diversity orders of paired NOMA users are zero.
  \item  We derive exact analytical expressions for ergodic rate in both LoS and NLoS scenarios. We provide tractable analytical lower bounds for the general case. We also provide closed-form expressions for the special case when the path loss is three. We obtain high SNR slopes for the paired NOMA users based on the developed ergodic rate. The obtained results confirm that the high SNR slopes of far users are zero in both LoS and NLoS scenarios.
  \item  We demonstrate that 1) the outage performance and ergodic rate can be enhanced by the LoS propagation; 2) the ergodic rates of near users have the same rate ceiling in both LoS and NLoS scenarios; 3) diversity orders and high SNR slopes of paired NOMA users are not affected by the LoS transmission of the proposed framework.
\end{itemize}

%Interestingly, in the UAV network with interference sources, the outage probability of the near user is much higher than the far user in most cases, which is different from previous literature.
\subsection{Organization and Notations}

The rest of the paper is organized as follows. In Section \uppercase\expandafter{\romannumeral2}, a model of UAV-aided transmission scenario is investigated in wireless networks, where NOMA users are uniformly allocated on the ground. Then precoding and detection strategies have proposed for UAV networks. Analytical results are presented in Section \uppercase\expandafter{\romannumeral3} to show the performance of UAV-aided MIMO-NOMA networks. Our numerical results are demonstrated in Section \uppercase\expandafter{\romannumeral4} for verifying our analysis, which is followed by the conclusion in Section \uppercase\expandafter{\romannumeral5}. Table~\ref{TABLE OF NOTATINS} lists all notations used in this article.

\begin{table}[h]
\small
\centering
\caption{\small{TABLE OF NOTATIONS}}
\begin{tabular}{|c|c|}
\hline
$\alpha _{{k}}^2$ and $\alpha _{{k'}}^2$ & Power allocation factors \\
\hline
$P_u$& Transmit power of the UAV\\
\hline
$P_I$& Transmit power of interference sources\\
\hline
${\rm \bf t}_k$ and ${\rm \bf t}_{k'}$& Detection vectors\\
\hline
$\alpha$&Path loss exponent\\
\hline
$R_m$ and $R_d$  & The radius of small disc and large disc \\
\hline
$R_k$ and $R_{k'}$  & The target rate of the $k$-th user and of the $k'$-th user\\
\hline
$m$ &Small-scale fading parameter\\
\hline
\end{tabular}
		
\label{TABLE OF NOTATINS}
\end{table}

\vspace{-0.1in}
\section{MIMO-NOMA Assisted UAV Network Model}
\begin{figure*}[t!]
\centering
\includegraphics[width =4in]{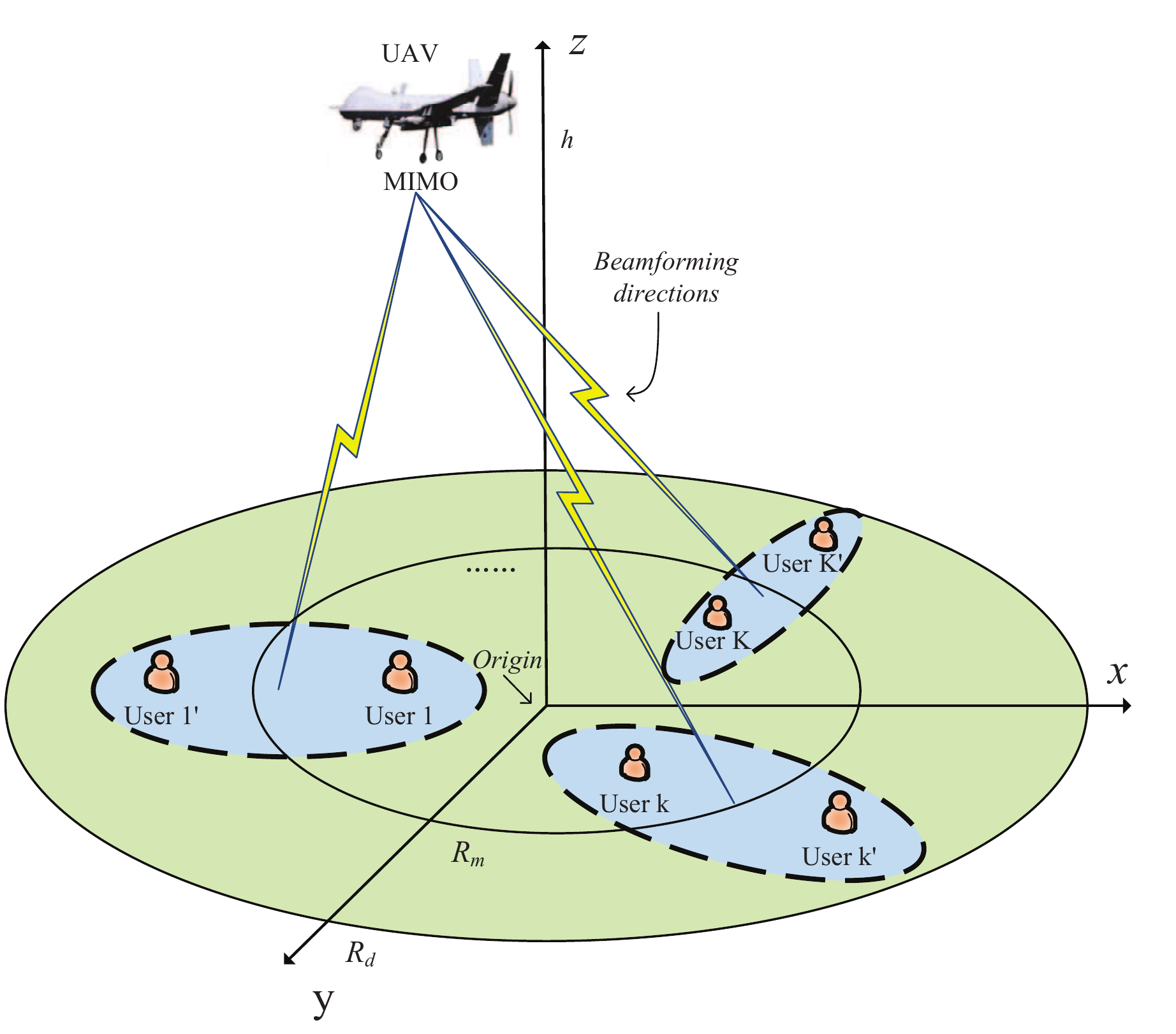}
\caption{Illustration of a typical UAV cellular network supported by beamforming.}
\label{system_model}
\vspace{-0.25in}
\end{figure*}

%\vspace{-0.02in}
Consider a MIMO-NOMA downlink communication scenario in which a UAV equipped with $K$ antennas is communicating with multiple users equipped with $N$ antennas each. Fig.~\ref{system_model} illustrates the wireless communication model with a single UAV, which is supported by MIMO beamforming, namely the cluster-based MIMO-NOMA. Multiple users are grouped into one cluster, and the UAV serves multiple clusters simultaneously in the cluster-based MIMO-NOMA, which can perfectly improve the system performance \cite{Liu_MIMO,Wang_QoE_perspective}.

\subsection{System Description}

The UAV cell coverage is a disc area, denoted by $\mathcal{D}$, which has a coverage radius of $R_d$. It is assumed that the users are uniformly distributed according to HPPPs distribution, which is denoted by $\Psi$ and associated with the density $\lambda$, within large disc $\mathcal{D}$ and small disc with radius $R_d$ and $R_m$, respectively. For simplicity, we only focus our attention on investigating a typical user pairing in this treatise, where two users are grouped to deploy NOMA transmission protocol.

The downlink users also detect signals sent by interferers which are distributed in $\mathbb{R}$ according to HPPPs distribution $\Psi_I$ with density $\lambda_I$ \cite{Stochastic_Geo}. It is assumed that the interference sources are equipped with one antenna each and use identical transmission power, which is denoted by $P_I$.

\subsection{Channel Model}

Consider the use of a composite channel model with two parts, large-scale fading and small-scale fading. $g_{k,kn}$ denotes the channel coefficient between the $k$-th antenna of the UAV and the $n$-th antenna of the $k$-th user, and the channel gain ${g_{k,kn}}$ is modelled as
\begin{equation}\label{channel gain,eq1}
{g_{k,kn}} = \beta _{k,kn} {h_{k,kn}},
\end{equation}
%\sqrt {\beta _{m,kn}^{ - \alpha }}
where $\beta _{k,kn}$ and $h_{k,kn}$ denotes the large-scale fading and small-scale fading, respectively. It is assumed that $\beta _{k,kn}$ and $h_{k,kn}$ with $k=1,... K, n=1,..., N$, are independent and identically distributed (i.i.d.). In this paper, large-scale fading represents the path loss and shadowing between the UAV and users. Generally speaking, the large-scale fading between the $k$-th user and the UAV changes slightly for different antenna pairs, which can be expressed as
\begin{equation}\label{large scale fading,eq2}
{\beta _k} = {\beta _{k,kn}},\forall k = 1 \cdots K,n = 1 \cdots N.
\end{equation}
For simplicity, the small-scale fading matrix from the UAV to user $k$ is defined as
\begin{equation}\label{channel matrix,eq3}
{\rm \bf H}_k = \left[ {\begin{array}{*{20}{c}}
{{h_{k,11}}}& \cdots &{{h_{k,K1}}}\\
 \vdots & \vdots & \vdots \\
{{h_{k,1N}}}& \cdots &{{h_{k,KN}}}
\end{array}} \right],
\end{equation}
where ${\rm \bf H}_k$ is a $N \times K$ matrix whose elements represent Nakagami-\emph{m} fading channel gains. The density function of the elements is ${f}(x) = \frac{{{m}^{{m}}{x^{{m} - 1}}}}{{\Gamma ({m})}}{e^{ - {{{m}x}}}}$, where $m$ denotes the fading parameter. In this paper, the UAV can be projected to the coverage disc by projection theorem. Thus, the distance between the UAV to user $k$ can be written as
\begin{equation}\label{projective_distance,eq4}
{d _{{k}}} = \sqrt {{h ^2} + {r_k^2}},
\end{equation}
where $h$ denotes the height of the UAV, and $r_k$ is the horizontal distance between the $k$-th user and projective point of the UAV. Therefore, the large-scale fading can be expressed as
\begin{equation}\label{large-scale fading}
\beta _k=\sqrt{{d _{k}^{ - \alpha }}} ,
\end{equation}
where $\alpha$ denotes the path loss exponent. For simplicity, we still use ${\beta _{{k}}}$ to represent the large-scale fading between the UAV and the $k$-th user in Section \uppercase\expandafter{\romannumeral2}.
Thus, the received power for the $k$-th user from the UAV is given by
\begin{equation}\label{received user power}
{P_{k}} = {P_u}{\beta _{k}}{\left| {{{\rm \bf H}_{k}}} \right|^2},
\end{equation}
where $P_u$ denotes the transmit power of the UAV. Besides, it is assumed that the CSI of users is perfectly known at the UAV. The proposed MIMO-NOMA assisted UAV network model for downlink transmission is described in the following subsection.

\subsection{Downlink transmission}
In this subsection, we estimate the downlink quality of MIMO-NOMA assisted single UAV networks, where two users, i.e., the $k$-th user and the ${k'}$-th user, are grouped to perform NOMA. We consider a more practicable method that the number of antenna equipped on each user is assumed to greater than $0.5K$, i.e., $N \ge 0.5K$.
In the beginning, it is assumed that the maximum clusters is $K$. Therefore, the UAV sends the following $K \times 1$ information-bearing vector $\rm \bf s$ to users as follows:
\begin{equation}\label{information-bearing s,eq5}
{\rm \bf{s}} = \left[ {\begin{array}{*{20}{c}}
{{\alpha _1}{s_1} + {\alpha _{1'}}{s_{1'}}}\\
 \vdots \\
{{\alpha _k}{s_k} + {\alpha _{k'}}{s_{k'}}}\\
 \vdots \\
{{\alpha _K}{s_K} + {\alpha _{K'}}{s_{K'}}}
\end{array}} \right],
\end{equation}
where $s_k$ is the signal intended for the $k$-th user, $\alpha_k$ is the power allocation coefficient, and $\alpha _k^2+\alpha _{k'}^2=1$.
The co-channel interference ${\rm \bf I}_{{k}}$ can be further expressed as follows:
\begin{equation}\label{interference,eq7}
{{\rm \bf I}_{{k}}} \buildrel \Delta \over = \sum\limits_{j \in \Psi_I } {\sqrt {{P _I}d_{j,k}^{ - \alpha }} } {{\bf 1}_{{N} \times {1}}},
\end{equation}
where ${\bf 1}_{{k} \times {n}}$ denotes an $k \times n$ all one matrix, $d_{j,k}$ denotes the distance from the $k$-th user to the $j$-th interference source. $\Psi_I$ and $P_I$ are the HPPPs distribution and the single antenna transmission power of interference sources, respectively. The interference model omits the small scale fading, since the effect of path loss is the dominant part for the long distance interferences.

Similar to~\cite{Ding_max_muller,MIMO_journal_DING_alignment}, the detection vectors $\rm \bf t$ can be designed as follows:
\begin{equation}\label{detection vector design}
\left[ {\begin{array}{*{20}{c}}
{{{\rm \bf{t}}_k}}\\
{{{\rm \bf{t}}_{k'}}}
\end{array}} \right] = {{\rm \bf{D}}_k}{{\rm \bf{x}}_k},
\end{equation}
where ${{\rm \bf{x}}_k}$ denotes a $(2N-K) \times 1$ vector, which is normalized to 2, i.e., ${\left| {{{\rm \bf{x}}_k}} \right|^2} = 2$. ${{\rm \bf{D}}_k}$ denotes a $2N \times (2N-K)$ matrix containing the $(2N-K)$ right singular vectors of $\left[ {\begin{array}{*{20}{c}}
{{\rm \bf{H}}_k^H}&{{\rm \bf{ - H}}_{k'}^H}
\end{array}} \right]$ corresponding to its zero singular values.

Define ${{\rm \bf b}_k} \buildrel \Delta \over = {\rm \bf{H}}_k^H{{\rm \bf{t}}_k}$ as the effective channel vector shared by users in the cluster, ${\rm \bf B} \buildrel \Delta \over = \left[ {\begin{array}{*{20}{c}}
{{\rm \bf b}_1}& \cdots &{{\rm \bf b}_K}
\end{array}} \right]^{H}$, and ${(\cdot)_{k,k}}$ denotes the $k$-th element on the main diagonal of the matrix.
Thus, with the design of precoding matrix $\rm \bf P$ and detection vector $\rm \bf t$ in~\cite{Ding_max_muller,MIMO_journal_DING_alignment}, the signal-to-interference-plus-noise ratio (SINR) of two users can be derived as
\begin{equation}\label{SINR_m',eq18}
SIN{R_{k'}} = \frac{{\rho {{\left| {{u_{k'}}} \right|}^2}\alpha _{{k'}}^2}}{{{{\left| {{{\rm \bf{t}}_{k'}}} \right|}^2} + \rho {{\left| {{u_{k'}}} \right|}^2}\alpha _k^2 + {{\left| {{\rm \bf{t}}_{k'}^H{{\bf 1}_{N \times 1}}} \right|}^2}\left( {\sum\limits_{{{j}} \in \Psi_I } {{P_I} {d}_{j,k'}^{ - \alpha }}}\right)}},
\end{equation}
and
\begin{equation}\label{SINR_m,eq19}
SIN{R_k} = \frac{{\rho {{\left| {{u_k}} \right|}^2}\alpha _{k}^2}}{{{{\left| {{{\rm \bf{t}}_k}} \right|}^2} +{{\left| {{\rm \bf{t}}_{k}^H{{\bf 1}_{N \times 1}}} \right|}^2}\left( {\sum\limits_{{{j}} \in \Psi_I } {{P_I}  {d}_{j,k}^{ - \alpha }}}\right)}},
\end{equation}
where $\rho$ is the transmit SNR, ${u_k} = \frac{{{\beta _k}}}{{{{({{\rm \bf {B}}^{ - 1}}{{\rm \bf {B}}^{ - H}})}_{k,k}}}}$ and ${u_{k'}} = \frac{{{\beta _{k'}}}}{{{{({{\rm \bf {B}}^{ - 1}}{{\rm \bf {B}}^{ - H}})}_{k',k'}}}}$. Again, it is assumed that the $k$-th user is the user with higher channel gain. Thus, based on NOMA protocol, the $k'$-th user treats the signal of user $k$ as noise, and the $k$-th user can neglect the signal from $k'$-th user by SIC technique. The use of signal alignment strategy reforms small-scale fading gains of two users into one matrix. Hence, it can be considered that the $k$-th user and the $k'$-th user are sharing the same small-scale fading matrix.
\vspace{-0.1in}
\section{Performance Evaluations}
In this section, we discuss the performance of downlink MIMO-NOMA assisted UAV networks. In this paper, fixed power allocation is employed at the UAV. New channel statistics, outage probabilities and ergodic rates are illustrated in the following three subsections.
\subsection{New Channel Statistics}
In this subsection, we derive new channel statistics for MIMO-NOMA assisted UAV networks, which will be used for estimating the outage probabilities and ergodic rates in the following subsections.

\begin{lemma}\label{lemma1:new channel state}
\emph{Assuming 2$K$ randomly located NOMA users in the disc of Fig.~\ref{system_model}, the detection vector of signal alignment assisted UAV networks over Nakagami-\emph{m} fading channels is }
\begin{equation}\label{detection vector result}
{{\rm \bf {D}}_k}{{\rm \bf {x}}_k}  \sim {\mathcal{CN}} \left( {0,m {\rm \bf I}_K} \right),
\end{equation}
\emph{where $m$ is the fading parameter of Nakagami-\emph{m} fading channels between the NOMA user and the UAV.}
\begin{proof}
Please refer to Appendix A.
\end{proof}
\end{lemma}

\begin{remark}\label{remark:channel state}
The result in~\eqref{detection vector result} indicates that the channel coefficients for the considered user pair are determined by the fading parameter of Nakagami-m fading channels.
\end{remark}
\textbf{Remark~\ref{remark:channel state}} provides insightful guidelines for estimating the outage performance of the proposed frameworks. The paired users share the same small scale fading channel coefficient, which follows exponential distribution with variance $m$, the performance of NOMA user pair can be calculated by their stochastic geometric approach.
\vspace{-0.1in}
\subsection{Outage Probabilities}
In this subsection, we first focus on the outage behavior of the far user $k'$, who is the user with poorer channel gain. The fixed power allocation strategy is deployed at the UAV, which the power allocation factors $\alpha_k$ and $\alpha_{k'}$ are constant during transmission. It is assumed that the target rate of user $k'$ is $R_{k'}$. Therefore, the outage probability of the $k'$-th user is given by
\begin{equation}\label{Outage}
{\rm P}_{k'}^O = {\rm P} \left( {\log \left( {1 + \frac{{\rho {{\left| {{u_{k'}}} \right|}^2}\alpha _{{k'}}^2}}{{{{\left| {{{\rm \bf{t}}_{k'}}} \right|}^2} + \rho {{\left| {{u_{k'}}} \right|}^2}\alpha _k^2 + {{\left| {{\rm \bf{t}}_{k'}^H{{\bf 1}_{N \times 1}}} \right|}^2}\sum\limits_{{{j}} \in {\Psi_I} } {{P_I}d_{j,k'}^{ - \alpha }} }}} \right) < {R_{k'}}} \right).
\end{equation}

The detection vector $\rm \bf t_{k'}$ and modified channel factor ${{u_{k'}}}$ are difficult to evaluate, which renders the evaluation of the outage probability very challenging. Therefore, we assume that ${{\left| {{{\rm \bf{t}}_{k'}}} \right|}^2}$ is 2, which is the greatest value under the constraint of~\eqref{detection vector design}. In addition, ${\left| {{\rm \bf {t}}_{k'}^H{{\bf 1}_{N \times 1}}} \right|^2}{\rm{ = }}{\left( {\sum\limits_{n = 1}^N {1 \cdot {t_n}} } \right)^2} \le {{N}}{\left| {{\rm \bf {t}}_{k'}^H} \right|^2}$. Hence, the outage probability of the $k'$-th user can be transformed into
\begin{equation}\label{transformed Outage}
{\rm \hat P}_{k'}^O = {\rm P}\left( {\log \left( {1 + \frac{{\rho {{\left| {{u_{k'}}} \right|}^2}\alpha _{{k'}}^2}}{{2 + \rho {{\left| {{u_{k'}}} \right|}^2}\alpha _k^2 + 2\delta \sum\limits_{{{j}} \in {\Psi _I}} {{P_I}d_{j,k'}^{ - \alpha }} }}} \right) < {R_{k'}}} \right),
\end{equation}
where $1 \le \delta  \le N$. Therefore, ${\rm \hat P}_{k'}^O$ is the upper bounds of the outage probability ${\rm P}_{k'}^O$ for $\delta = N$ and ${{{\left| {{{\rm \bf {t}}_{k'}}} \right|}^2}}=2$.

\noindent \textbf{Constraint 1} (Target rate). The target rate of the $k'$-th user satisfies $\rho \alpha _{{k'}}^2 - \rho \alpha _{k}^2{\varepsilon _{k'}} > 0$, where ${\varepsilon _{k'}} = {2^{{R_{k'}}}} - 1$. Otherwise, the outage probabilities of paired NOMA users are one.

\textbf{Constraint 1} provides the basic guideline of parameter setting for MIMO-NOMA assisted UAV networks with fixed power allocation. In NOMA, the far user $k'$ treats the signal of the $k$-th user as noise, and decodes the signal directly by SIC technique. Thus, the outage probability of the $k'$-th user is one unless the target rate of the far user must under the constraint ${R_{k'}} < {\log _2}\left( {1 + \frac{{\alpha _{k'}^2}}{{\alpha _k^2}}} \right)$. On the other hand, the near user needs to decode the signal intended for the far user before decoding its own signal. Thus, \textbf{Constraint 1} is needed to satisfy for both NOMA users.

Then we turn our attention to calculating the outage probability of the far user with interference source, which is given in the following Theorem.
\begin{theorem}\label{Theorem1:Outage m' upper}
\emph{Given any random vector ${\rm \bf x}_{k'}$, the upper bound of the outage probability for the far user can be expressed as}
\begin{equation}\label{outage uper bounds in theorem1}
{\rm{\hat P}}_{_{k'}}^O = {\rm{1}} - \frac{2}{{R_d^2 - R_m^2}}\int_{{l_{{m}}} }^{{l_{{d}}} } { {\exp \left( { - {\frac{2}{m}}{V_{k'}}{{x}}_{}^\alpha } \right)\exp \left( { - {\lambda _I}\pi \phi _{k'}^{\frac{2}{\alpha }}\gamma \left( {\begin{array}{*{20}{c}}
{\frac{1}{\alpha },}&{\frac{{{\phi _{k'}}}}{{r_0^\alpha }}}
\end{array}} \right)} \right)  xdx}  },
\end{equation}
\emph{where ${V_{k'}} = \frac{{{\varepsilon _{k'}}}}{{{\rho \alpha _{{k'}}^2 - \rho \alpha _{k}^2{\varepsilon _{k'}}} }}$, ${\phi _{k'}} = {\frac{2}{m}}{V_{k'}}\delta {P_I}x^\alpha$, ${l_{{d}}} = \sqrt {{h^2}+ R_d^2 }$, ${l_{{m}}} = \sqrt {{h^2} +R_m^2 }$, and $\gamma \left(  \cdot  \right)$ represents the lower incomplete Gamma function.}
\begin{proof}
Please refer to Appendix B.
\end{proof}
\end{theorem}

In the proposed framework, the lower bound of the outage probability, which is meaningless, is only affected by power allocation factors due to that the minimum detection vector approaches zero. Thus, in this article, the lower bound of outage probability will not be estimated.

It is challenging to solve the integral in~\eqref{outage uper bounds in theorem1} directly due to the lower incomplete Gamma function. Thus, in order to derive the diversity order and gain further insights into the system's operation in the high SNR regime, the asymptotic behavior is analyzed, usually when the SNR of the channels between the UAV and users is sufficiently high, i.e., when the transmit SNR obeys ${\rho  \to \infty }$.

\begin{corollary}\label{Corollary1: m' asymptotic}
\emph{ Assuming that $\rho \alpha _{{k'}}^2 - \rho \alpha _{k}^2{\varepsilon _{k'}} > 0$, the asymptotic outage probability of the $k'$-th user is given by}
\begin{equation}\label{asympto m'}
{\rm{\hat P}}_{_{k'}}^\infty  = \frac{{2{T_{k'}}}}{{R_d^2 - R_m^2}}\frac{{{{({h^2} + R_d^2)}^{\frac{{\alpha  + 2}}{2}}} - {{({h ^2} + R_m^2)}^{\frac{{\alpha  + 2}}{2}}}}}{{\alpha  + 2}},
\end{equation}
\emph{where ${T_{k'}} = {\frac{2}{m}}{V_{k'}}\left( {1 + \pi {\lambda _I}\delta {P_I}\frac{\alpha }{{{r_0}}}} \right)$.}
\begin{proof}
Please refer to Appendix C.
\end{proof}
\end{corollary}

\begin{remark}\label{remark:asymptotic m'}
The derived results in \eqref{asympto m'} demonstrate that the outage probability of the far user can be decreased by increasing the fading parameter $m$ or decreasing the target rate of the far user itself.
\end{remark}

\begin{proposition}\label{proposition1: m' diversity order}
\emph{From \textbf{Corollary \ref{Corollary1: m' asymptotic}}, one can yield the diversity order by using the high SNR approximation, and the diversity order of the $k'$-th user in the proposed MIMO-NOMA assisted UAV networks is given by}
\begin{equation}\label{diversity order}
{w_{k'}} =  - \mathop {\lim }\limits_{\rho  \to \infty } \frac{{\log {\rm{\hat P}}_{k'}^\infty }}{{\log \rho }} \approx 1.
\end{equation}
\end{proposition}

The exact expression in the case of $P_I=0$ is also worth calculating, which is given in the following corollary.

\begin{corollary}\label{Corollary2: m' exact, inter not existed}
\emph{For the special case that the interference sources does not exist, the exact expression can be written as}
\begin{equation}\label{non interference existed_k'}
\begin{aligned}
{\rm{\hat P}}_{{k'}}^O =& {\rm{1}} - \frac{1}{{R_d^2 - R_m^2}}\left( {l_{{d}}^2{e^{{\rm{ - }}{\frac{2}{m}}{V_{k'}}l_{{d}}^\alpha }} - l_{{m}}^2{e^{{\rm{ - }}{\frac{2}{m}}{V_{k'}}l_{{m}}^\alpha }}} \right) \\
&- \frac{{{{\left( {{\frac{2}{m}}{V_{k'}}} \right)}^{ - \frac{2}{\alpha }}}}}{{R_d^2 - R_m^2}}\left( {\gamma \left( {\frac{2}{\alpha } + 1,{\frac{2}{m}}{V_{k'}}l_{{d}}^\alpha } \right) - \gamma \left( {\frac{2}{\alpha } + 1,{\frac{2}{m}}{V_{k'}}l_{{m}}^\alpha } \right)} \right).
\end{aligned}
\end{equation}
\begin{proof}
In the special case of ${\rm \mathbb E}\left\{ {{d_{j,k'}}} \right\}=1$, the outage probability for the far user can be rewritten as
\begin{equation}\label{outage m' exact interfer not}
{\rm{\hat P}}_{_{k'}}^O = {\rm{1}} - \frac{1}{{R_d^2 - R_m^2}}\int_{{l_{{m}}} }^{{l_{{d}}} } { {\exp \left( { - {\frac{2}{m}}{V_{k'}}{{x}}_{}^\alpha } \right)d\left(x^2 \right)} }.
\end{equation}
Considering the partial integration for \eqref{outage m' exact interfer not}, the exact expression can be converted as
\begin{equation}\label{results of the exact solution}
{\rm{\hat P}}_{{k'}}^O= 1 - \frac{1}{{R_d^2 - R_m^2}}\left( {l_d^2\exp \left( { - \frac{2}{m}{V_{k'}}l_d^\alpha } \right) - l_m^2\exp \left( { - \frac{2}{m}{V_{k'}}l_m^\alpha } \right)
- \int_{{l_m}}^{{l_d}}  {{x^2}d\left( \exp \left( { - \frac{2}{m}{V_{k'}}x_{}^\alpha } \right) \right)} } \right).
\end{equation}
Thus, Corollary~\ref{Corollary2: m' exact, inter not existed} is proved.
\end{proof}
\end{corollary}

In NOMA, the $k$-th user needs to decode the signal intended for user $k'$ before decoding its own signal via SIC technique. Therefore, the outage probability of the $k$-th user is given by
\begin{equation}\label{outage for m}
\begin{aligned}
{\rm{P}}_k^O &= {\rm{P}}\left( {\log \left( {1 + \frac{{\rho {{\left| {{u_k}} \right|}^2}\alpha _{{k'}}^2}}{{{{\left| {{{\rm \bf {t}}_k}} \right|}^2} + \rho {{\left| {{u_k}} \right|}^2}\alpha _k^2 + {{\left| {{\rm \bf {t}}_{k}^H{{\bf 1}_{N1}}} \right|}^2}\sum\limits_{{j} \in {\Psi _I}} {{P_I}d_{j,k}^{ - \alpha }} }}} \right) < {R_{k'}}} \right) \\
& + {\rm{P}}\left( {\log \left( {1 + \frac{{\rho {{\left| {{u_k}} \right|}^2}\alpha _{{k'}}^2}}{{{{\left| {{{\rm \bf {t}}_k}} \right|}^2} + \rho {{\left| {{u_k}} \right|}^2}\alpha _k^2 + {{\left| {{\rm \bf {t}}_{k}^H{{\bf 1}_{N1}}} \right|}^2}\sum\limits_{{j} \in {\Psi _I}} {{P_I}d_{j,k}^{ - \alpha }} }}} \right) > {R_{k'}},}\right. \\
& \ \ \left. {\log \left( {1 + \frac{{\rho {{\left| {{u_k}} \right|}^2}\alpha _{{k}}^2}}{{{{\left| {{{\rm \bf {t}}_k}} \right|}^2} + {{\left| {{\rm \bf {t}}_{k}^H{{\bf 1}_{N1}}} \right|}^2}\sum\limits_{{j} \in {\Psi _I}} {{P_I}d_{j,k}^{ - \alpha }} }}} \right) < {R_k}} \right),
\end{aligned}
\end{equation}
where ${R_k}$ denotes the target rate of the near user $k$.
Again, we focus on the upper bound of the outage probability for user $k$. By some algebraic handling, the upper bound of the $k$-th user's outage probability can be rewritten as
\begin{equation}\label{outage upper expression}
{\rm{\hat P}}_k^O = {\rm{P}}\left( {{{\left| {{u_k}} \right|}^2} < {\frac{2}{m}}{V_{k'}}{\varphi _k}} \right) + {\rm{P}}\left( {{\frac{2}{m}}{V_{k'}}{\varphi _k} < {{\left| {{u_k}} \right|}^2} < {\frac{2}{m}}{V_k}{\varphi _k}} \right),
\end{equation}
where ${\varepsilon _k} = {2^{{R_k}}} - 1$, ${V_k} = \frac{{{\varepsilon _k}}}{{ \rho \alpha _k^2}}$, and ${\varphi _k} = 1 + \delta {P_I}\sum\limits_{{j} \in {\Psi _I}} {d_{j,k}^{ - \alpha }}$. The expression of outage probability for the near user is provided in the following theorem.

\begin{theorem}\label{Theorem2:Outage m upper}
\emph{Under \textbf{Constraint} 1, the upper bound of outage probability ${\rm{\hat P}}_k^O$ for the near user can be written as}
\begin{equation}\label{outage m}
{\rm{\hat P}}_{k}^O = {\rm{1}} - \frac{2}{{R_m^2}}\int_h ^{{l_{{m}}} } {\left( {\exp \left( { - {\frac{2}{m}}{\max \left\{ {{V_{k'}},{V_k}} \right\}}x_{}^\alpha } \right)\exp \left( { - {\lambda _I}\pi \phi_{k}^{\frac{2}{\alpha }}\gamma \left( {\begin{array}{*{20}{c}}
{\frac{1}{\alpha },}&{\frac{{{\phi _{k}}}}{{r_0^\alpha }}}
\end{array}} \right)} \right)  } \right)}xdx,
\end{equation}
\emph{if $\rho \alpha _{{k'}}^2 - \rho \alpha _{k}^2{\varepsilon _{k'}} > 0$}

\begin{proof}
Please refer to Appendix D.
\end{proof}
\end{theorem}

Based on \textbf{Theorem}~\ref{Theorem2:Outage m upper}, we can calculate the asymptotic result of the near user in the following corollary.

\begin{corollary}\label{Corollary3: m asymptotic}
\emph{In the case where $\rho$ approaches infinity, it is straightforward to attain that ${\max \left\{ {{V_{k'}},{V_k}} \right\}}$ goes to zero. Therefore, the asymptotic result can be obtained as}
\begin{equation}\label{aysmptotic simulation m}
\begin{aligned}
{\rm{\hat P}}_{k}^\infty & =  \frac{{2{T_k}}}{{R_m^2}}\frac{{{{({h ^2} + R_m^2)}^{\frac{{\alpha  + 2}}{2}}} - {h ^{\alpha  + 2}}}}{{\alpha  + 2}},
\end{aligned}
\end{equation}
\emph{where ${T_k} = \frac{{2\max \left\{ {{V_{k'}},{V_k}} \right\}}}{m}\left( {1 + \pi {\lambda _I}\delta {P_I}\frac{\alpha }{{{r_0}}}} \right)$.}
\begin{proof}
Similar to Appendix C, the asymptotic outage probability of the near user can be proved.
\end{proof}
\end{corollary}

\begin{remark}\label{remark:diversity m}
Due to the fact that two users share the same small scale fading matrix by signal alignment technique, one can obtain that the diversity order of the $k$-th user is also one.
\end{remark}

\begin{proposition}\label{proposition2: m' diversity order}
\emph{For the case that the power level of interference sources increases as transmit SNR increases, one can easily yield the diversity order from \textbf{Corollary \ref{Corollary1: m' asymptotic}} and \textbf{Corollary \ref{Corollary3: m asymptotic}}, and the diversity order of the paired NOMA users is given by}
\begin{equation}\label{diversity order, dynamic power of interference}
{w_{k'}} = {w_{k}} \approx 0.
\end{equation}
\end{proposition}

\begin{remark}\label{remark:asymptotic m' large interference}
The diversity order in~\eqref{diversity order, dynamic power of interference} demonstrates that the power of interference sources, which is proportional to transmit SNR, influences the outage performance dramatically.
\end{remark}

\begin{corollary}\label{Corollary2: m exact, inter not existed}
\emph{For the special case that the interference is limited, the exact expression for the near user can be rewritten as}
\begin{equation}\label{non interference existed}
\begin{aligned}
&{\rm{\hat P}}_{{k}}^O = {\rm{1}} - \frac{1}{{R_m^2}}\left( {l_{{m}}^2{e^{{\rm{ - }}{\frac{2}{m}}\max \left\{ {{V_{k'}},{V_k}} \right\}l_{{m}}^\alpha }}  - h^2{e^{{\rm{ - }}{\frac{2}{m}}\max \left\{ {{V_{k'}},{V_k}} \right\}h^\alpha }}  } \right) \\
&- \frac{{{{\left( {{\frac{2}{m}}\max \left\{ {{V_{k'}},{V_k}} \right\}} \right)}^{ - \frac{2}{\alpha }}}}}{{R_m^2}}\left( {\gamma \left( {\frac{2}{\alpha } + 1,{\frac{2}{m}}\max \left\{ {{V_{k'}},{V_k}} \right\}l_{{m}}^\alpha  } \right) - \gamma \left( {\frac{2}{\alpha } + 1,{\frac{2}{m}}\max \left\{ {{V_{k'}},{V_k}} \right\}h^\alpha } \right)} \right).
\end{aligned}
\end{equation}
\end{corollary}
\vspace{-0.1in}
\subsection{Ergodic Rates}
The ergodic rate is a critical metric for performance evaluation. Therefore, in this subsection, we focus on analyzing the ergodic rates of individual users, which are determined by their channel conditions and geometry parameters in the proposed framework. The ergodic rates of the far user is given by
\begin{equation}\label{ergodic rate m' eq}
{R_{k'}} = \mathbb {E}\left\{ {{{\log }_2}\left( {1 + SIN{R_{k'}}\left( {{x_{k'}}} \right)} \right)} \right\}.
\end{equation}

Thus, the ergodic rate for the far user $k'$ can be obtained in the following theorem.

\begin{theorem}\label{Theorem3:m' ergodic rate}
\emph{The closed expression of ergodic rate for the $k'$-th user is given by}
\begin{equation}\label{ergodic asymp m' express in paper}
{R_{k'}} = {\log _2}(1 + \hat \alpha ) + {Q_{k'}}\hat \alpha {\log _2}(1 + {\alpha _{k}^2}) - {Q_{k'}}{\log _2}(1 + \hat \alpha ).
\end{equation}
\emph{where ${Q_{k'}} = \frac{{4\left(  {1 + \frac{\pi {\lambda _I}\delta {P_I} }{{{\alpha r_0}}}} \right)\left( {l_d^{\alpha  + 2} - l_m^{\alpha  + 2}} \right)}}{{m\rho \left( {R_d^2 - R_m^2} \right)\left( {\alpha  + 2} \right)}}$, and $\hat \alpha  \buildrel \Delta \over = \frac{{\alpha _{k'}^2}}{{\alpha _k^2}}$.}
\begin{proof}
Please refer to Appendix E.
\end{proof}
\end{theorem}

\begin{remark}\label{remark:ergodic m'}
One can observe from the result in~\eqref{ergodic asymp m' express in paper} that the ergodic rate of the far user is entirely dependent on the power allocation factors in the high SNR regime.
\end{remark}

The integration of lower incomplete gamma function is quit challenging to estimate. Thus, we only derive the ergodic rate of the near user in two special cases, i.e., high SNR approximation and exact expression in the case of fixed path loss exponent.

\begin{theorem}\label{Theorem4:m ergodic rate}
\emph{When transmit SNR obeys ${\rho  \to \infty }$, the approximation ergodic rate for the $k$-th user is given by}
\begin{equation}\label{m ergodic asym}
\begin{aligned}
{R_k} &= \frac{{{h^2}}}{{\ln \left( 2 \right)R_m^2}}{e^{{\rm{  }}C{h^\alpha }}}{\rm{Ei}}\left( -{C{h^\alpha }} \right) - \frac{{l_m^2}}{{\ln \left( 2 \right)R_m^2}}{e^{{\rm{ }}Cl_m^\alpha }}{\rm{Ei}}\left( - {Cl_m^\alpha } \right) \\
& +\sum\limits_{p = 0}^\infty  {\sum\limits_{i = 1}^{p + 1} {\frac{{{{\alpha \left( { - 1} \right)}^{i - 1}}\left( {i - 1} \right)!l_m^{\alpha p + \alpha  + 2 - \alpha i}{C^{p + 1 - i}}}}{{\left( {\frac{2}{\alpha } + 1} \right) \cdot  \ldots  \cdot \left( {\frac{2}{\alpha } + 1 + p} \right)\ln \left( 2 \right)R_m^2}}} }
- \sum\limits_{p = 0}^\infty  {\sum\limits_{i = 1}^{p + 1} {\frac{{{{\alpha  \left( { - 1} \right)}^{i - 1}}\left( {i - 1} \right)!h^{\alpha p + \alpha  + 2 - \alpha i}{C^{p + 1 - i}}}}{{\left( {\frac{2}{\alpha } + 1} \right) \cdot  \ldots  \cdot \left( {\frac{2}{\alpha } + 1 + p} \right)\ln \left( 2 \right)R_m^2}}} }
,
\end{aligned}
\end{equation}
\emph{where $C = \frac{{2\left(  {1 + \frac{\pi {\lambda _I}\delta {P_I} }{{{\alpha r_0}}}} \right)}}{{m\rho \alpha _{k}^2}}$.}
\begin{proof}
Please refer to Appendix F.
\end{proof}
\end{theorem}

The exact expression of the ergodic rate for the near user is difficult to calculate. Therefore, the special case, which path loss exponent is fixed, is shown in the following corollary.

\begin{corollary}\label{Corollary4: m ergodic when a=3}
\emph{In the special case in which path loss exponent $\alpha=3$, the exact expression of the ergodic rate for the near user can be attained to}
\begin{equation}\label{exact ergodic m}
\begin{aligned}
{R_k} &= \frac{{{h^2}}}{{\ln \left( 2 \right)R_m^2}}{e^{{\rm{  }}C{h^\alpha }}}{\rm{Ei}}\left( -{C{h^\alpha }} \right) - \frac{{l_m^2}}{{\ln \left( 2 \right)R_m^2}}{e^{{\rm{ }}Cl_m^\alpha }}{\rm{Ei}}\left( - {Cl_m^\alpha } \right)\\
&+\frac{{{C^{ -1}l_m^{- 1}}}}{{\ln \left( 2 \right)R_m^2}}G
\begin{tiny}
\begin{array}{*{20}{c}}
{2,2}\\
{3,2}
\end{array}
\end{tiny}
\left( \begin{array}{l}
-1,0,-\frac{2}{3}\\
-\frac{1}{3},0
\end{array} \bigg| \frac{1}{{Cl_m^3}} \right)
-\frac{{{C^{ -1}h^{- 1}}}}{{\ln \left( 2 \right)R_m^2}}G
\begin{tiny}
\begin{array}{*{20}{c}}
{2,2}\\
{3,2}
\end{array}
\end{tiny}
\left( \begin{array}{l}
-1,0,-\frac{2}{3}\\
-\frac{1}{3},0
\end{array} \bigg| \frac{1}{{Ch^3}} \right),
\end{aligned}
\end{equation}
\emph{where $G\left(  \cdot  \right)$ represents Meijer-G function.}
\begin{proof}
Please refer to Appendix G.
\end{proof}
\end{corollary}

\begin{remark}\label{remark:ergodic m}
One can know from the results in~\eqref{m ergodic asym} and \eqref{exact ergodic m} that the ergodic rate of the near user is mainly dependent on the transmit SNR.
\end{remark}
%$\mathbb{P}\mathbb{E}$

To gather deep insights for the system performance, the high SNR slope, as the key parameter determining the ergodic rate in high SNR regime, is worth estimating. Therefore, we first express the high SNR slope as
\begin{equation}\label{High SNR slope of m'}
S_\infty ^{k'} = \mathop {\lim }\limits_{x \to \infty } \frac{{{R_{k'}}}}{{{{\log }_2}\left( {1 + x} \right)}}.
\end{equation}

\begin{proposition}\label{proposition2: high SNR slope}
\emph{Substituting \eqref{ergodic asymp m' express in paper} into \eqref{High SNR slope of m'}, we obtain}
\begin{equation}\label{High SNR slope result of m'}
S_\infty ^{k'} = 0.
\end{equation}

\emph{Then plugging \eqref{m ergodic asym} into \eqref{High SNR slope of m'}, the high SNR slope of the near user can be illustrated as }
\begin{equation}\label{High SNR slope result of m}
S_\infty ^k = 1.
\end{equation}
\end{proposition}

\begin{remark}\label{remark:High SNR slope}
In the proposed framework, the ergodic rate for the far user can be consider as a constant in the high SNR regime from \textbf{Proposition~\ref{proposition2: high SNR slope}}.
\end{remark}
\vspace{-0.15in}
\section{Numerical Studies}
In this section, numerical results are provided to facilitate the performance evaluation of MIMO-NOMA assisted UAV networks. Mont Carol simulations are conducted to illustrate the correctness of analytical results. In the considered network, it is assumed that the power allocation factors are $\alpha_{k'}^2=0.75$ and $\alpha_{k}^2=0.25$. Some simulation parameters are summarized in Table~\ref{TABLE OF PARAMETERS}.
\vspace{-0.2in}
\begin{table}[h]
\small
\centering
\caption{\small{SIMULATION PARAMETERS}}
\begin{tabular}{|c|c|}
\hline
Mont Carlo Simulations repeated&$10^6$ times\\
\hline
The radius of the large plane&20 m\\
\hline
The radius of the small plane&10 m\\
\hline
The height of the UAV &10 m\\
\hline
Path loss exponent&$\alpha=3$\\
\hline
The interference sources density  & $\lambda_I= 10^{-4} $ \\
\hline
The radius of the interference sources &1000 m\\
\hline
The noise power  & $\sigma= 0 $ dBm\\
\hline
The antenna number of the UAV  &4\\
\hline
The antenna number of users &3\\
\hline
\end{tabular}
\vspace{-0.1in}
\label{TABLE OF PARAMETERS}
\end{table}

\vspace{-0.1in}
\subsection{Outage Probabilities}
\begin{figure*}[t!]
\centering
\subfigure[Outage probability of NOMA with interference sources versus transmit SNR in NLoS scenario.]{\label{Outage_individual with m=1}
\includegraphics[width =3.1in]{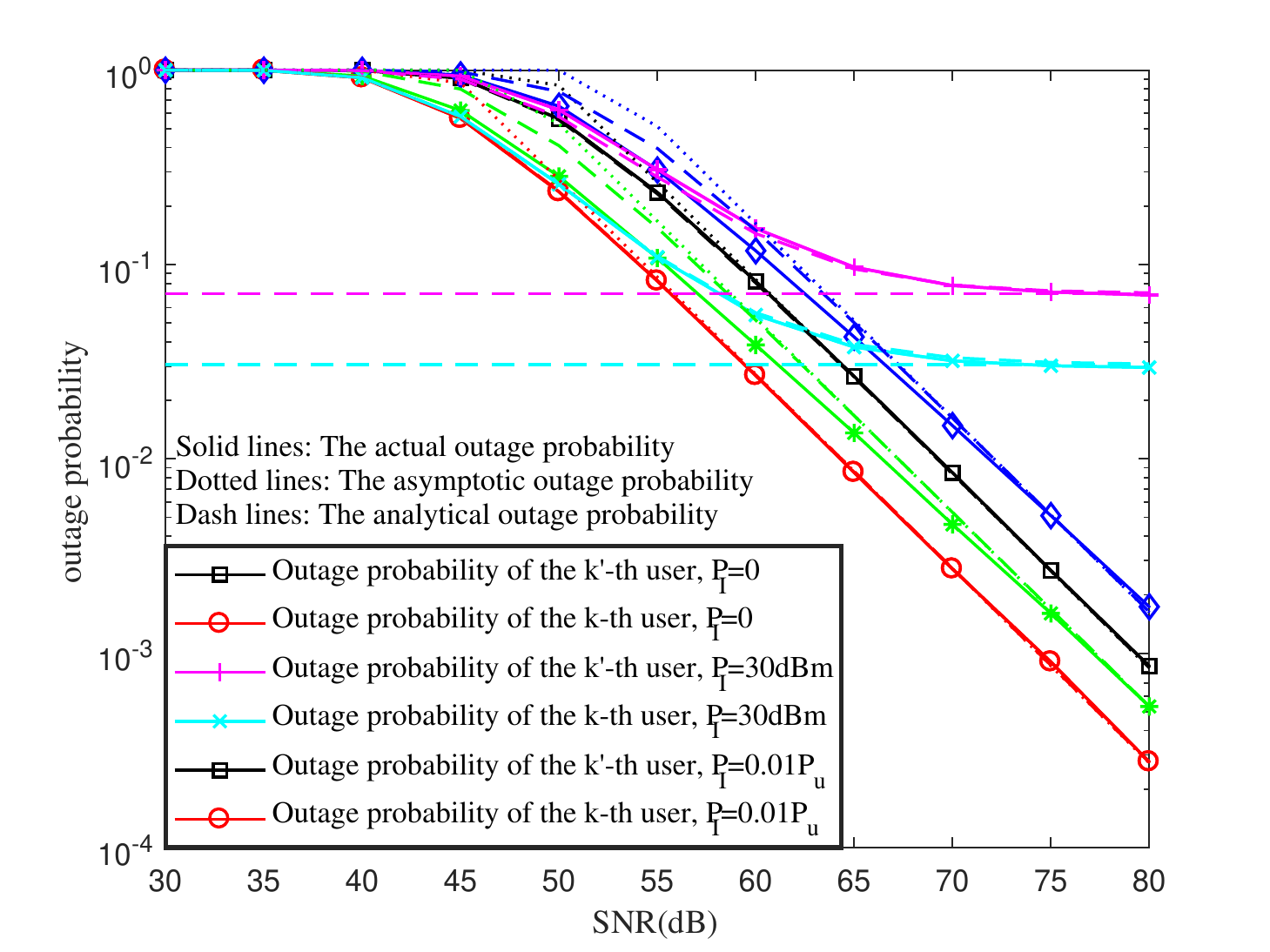}}
\subfigure[Outage probability of NOMA versus transmit SNR in LoS scenario.]{\label{Outage_individual with m=2}
\includegraphics[width= 3.1in]{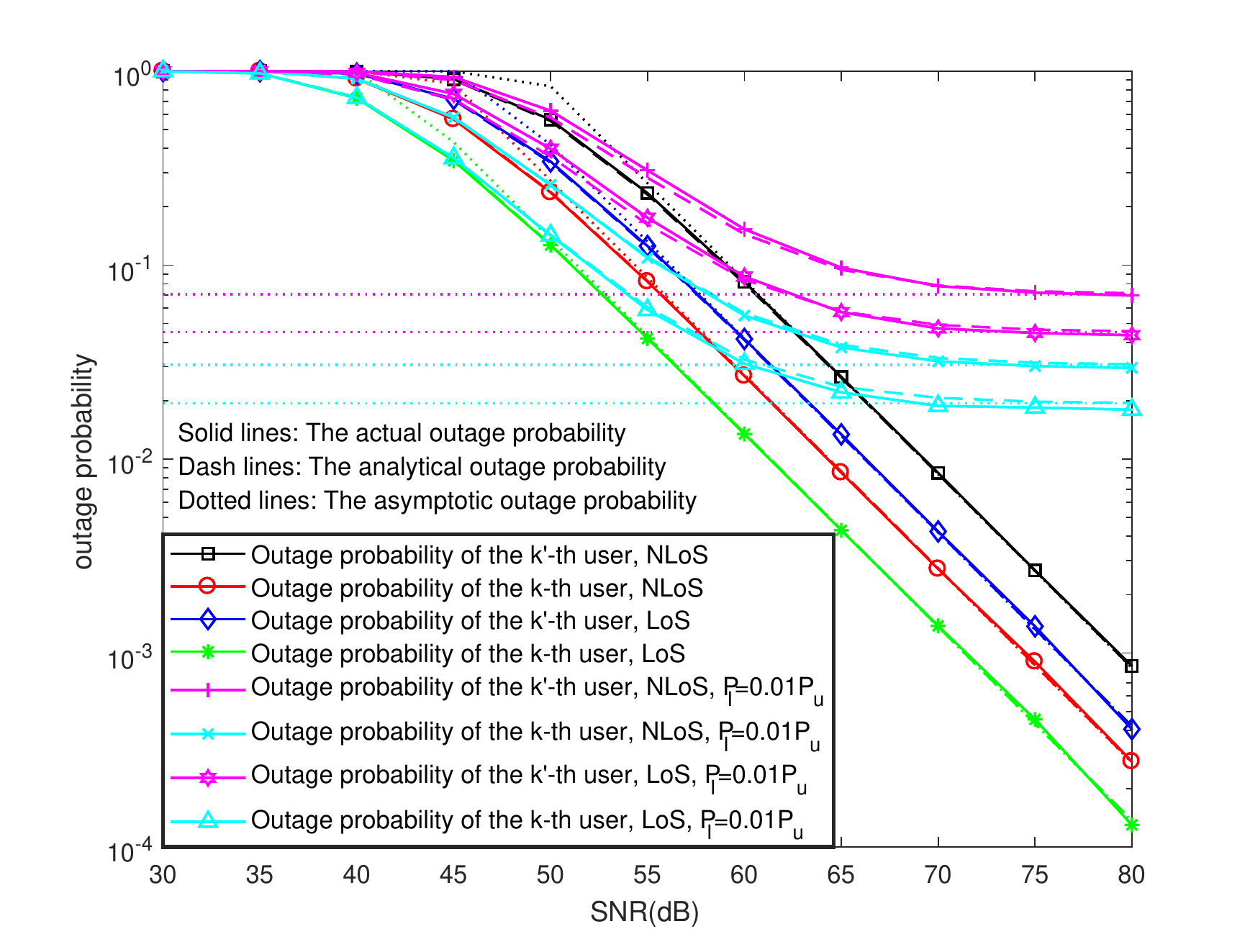}}
\caption{Outage probability of NOMA users versus transmit SNR, with target rate $R_{k'}=R_k=1.5$ bits per channel user (BPCU). The analytical results of NOMA are calculated from \eqref{non interference existed_k'} and \eqref{non interference existed}, and the asymptotic results are derived from \eqref{asympto m'} and \eqref{aysmptotic simulation m}.}
\label{Fig1:Outage_individual}
\vspace{-0.25in}
\end{figure*}

\begin{figure*}[t!]
\centering
\subfigure[Outage probability of NOMA with interference sources versus the height of the UAV and target rate, with $R_k=R_{k'}=R$.]{\label{m2_3D_outage}
\includegraphics[width =3.1in]{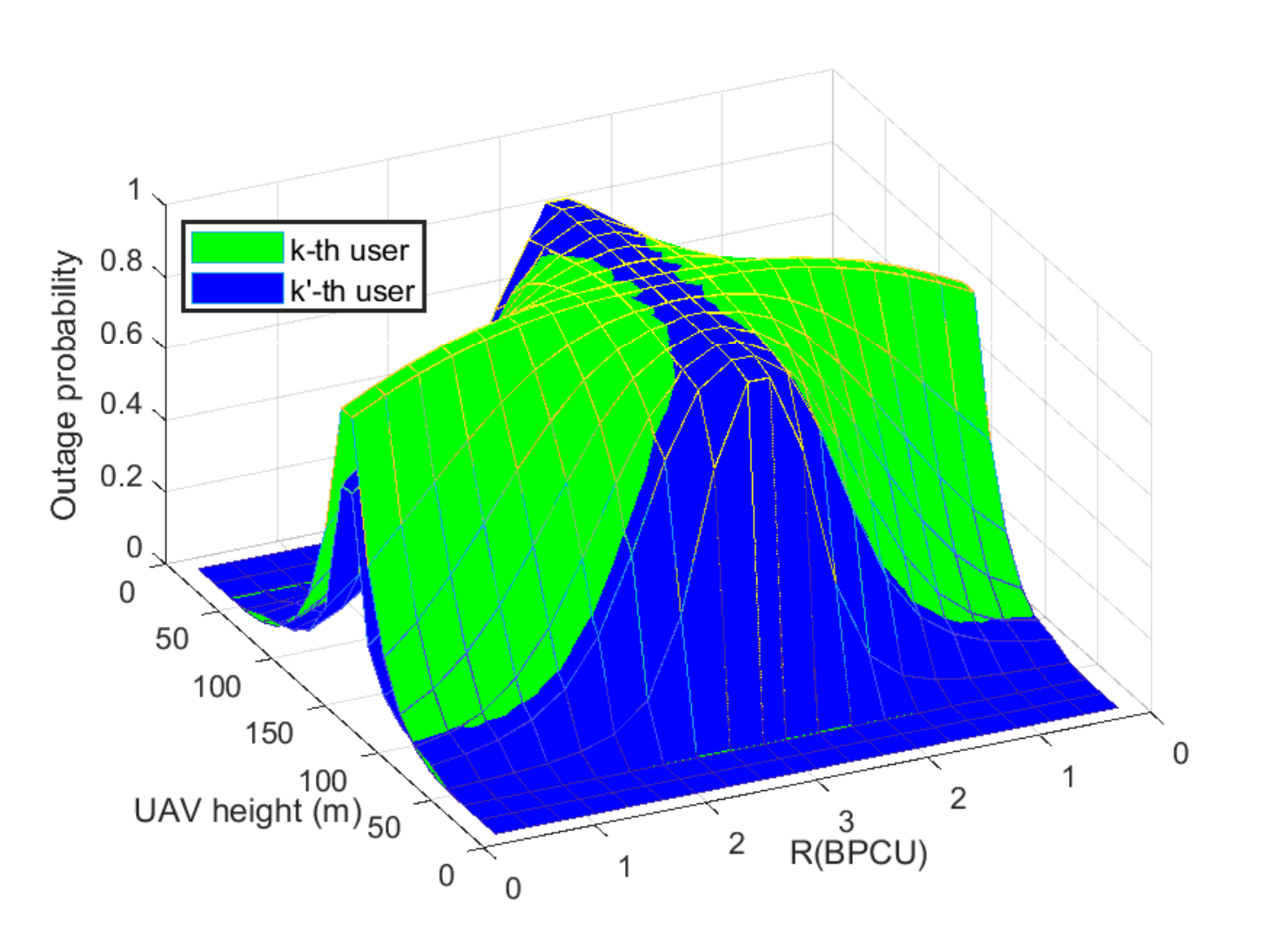}}
\subfigure[Outage probability of NOMA with interference sources versus the height of the UAV and radius, with $R_k=R_{k'}=1$BPCU.]{\label{m2_3D_outage_height_radius}
\includegraphics[width =3.1in]{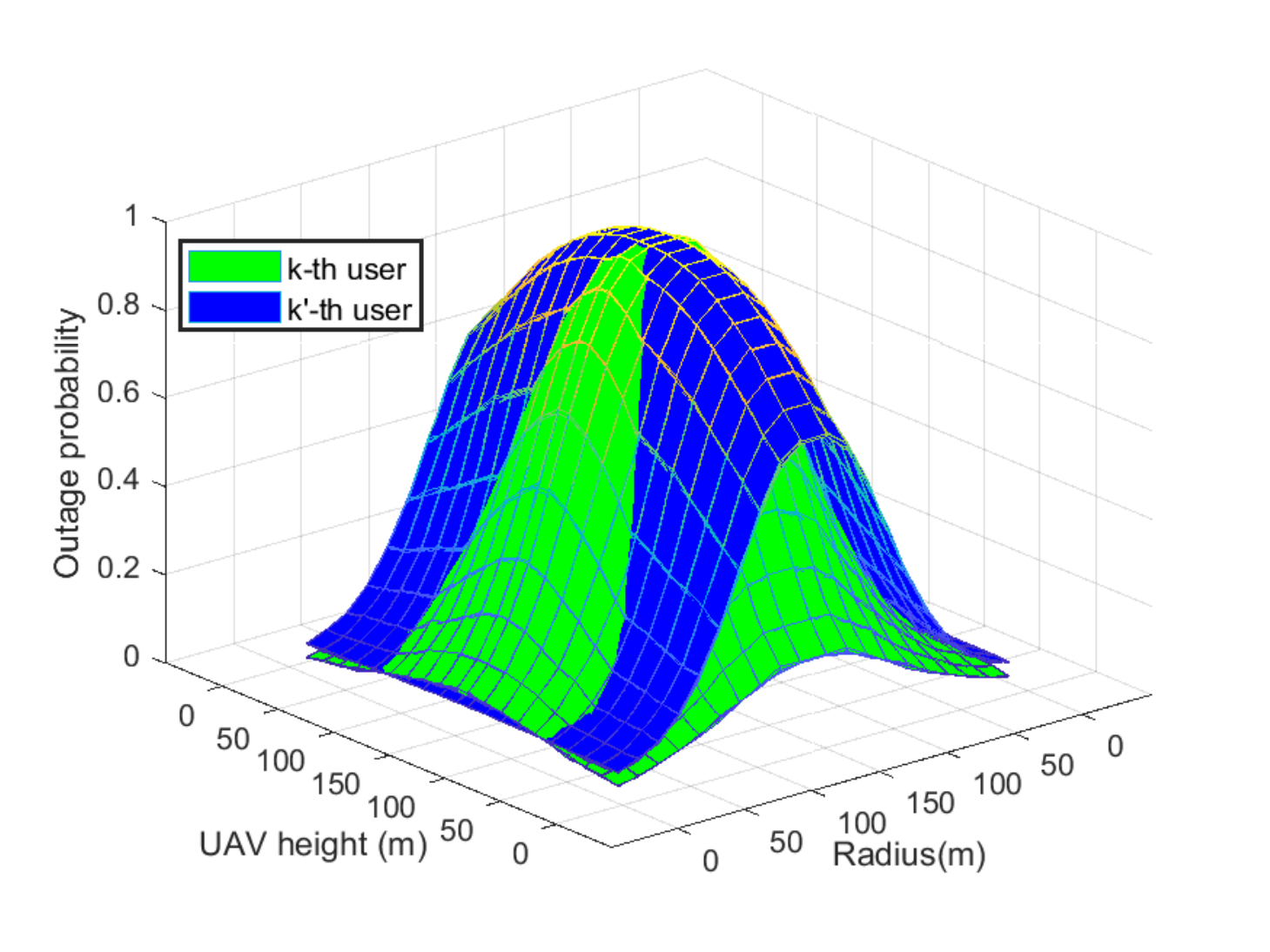}}
\label{Fig3:Outage_3D}
\caption{Outage probability of NOMA in LoS scenario, with $\rho=60$dB.}
\vspace{-0.25in}
\end{figure*}

In this subsection, the outage probability achieved by paired NOMA users with different choices of interference power in both LoS and NLoS scenarios is demonstrated in Fig. \ref{Outage_individual with m=1}, Fig. \ref{Outage_individual with m=2}, Fig. \ref{m2_3D_outage}, and Fig. \ref{m2_3D_outage_height_radius}.

In Fig.~\ref{Outage_individual with m=1}, we evaluate the performance of the proposed approach for a network with different interference power levels in NLoS scenario. In Fig.~\ref{Outage_individual with m=1}, the black and blue curves are the outage probability for far users, and the red and green curves are the outage probability for near users. We can see that, as interference power increases, the outage probability of both the NOMA users increases. This is due to the fact that, as higher power level of interference sources are deployed, the received SINR decreases. It is also confirmed the close agreement between the simulation and analytical results in high SNR regime for the fixed interference power scenario. Besides, it is worth noting that all curves in fixed interference power scenario have the same slopes, which indicates that the diversity orders of the schemes are all one. This phenomenon validates the insights from \textbf{Proposition \ref{proposition1: m' diversity order}} and \textbf{Remark \ref{remark:diversity m}}. Note that error floors appear when $P_I$ is proportional to the UAV power, which meet the expectation due to the strong co-channel interference. In this case, the dynamic interference power, which is a function of transmit SNR, is much greater than noise power in the high SNR regime. Thus, the error floor occurs in the case of high SNR situation. It is also worth noting that the asymptotic result does not exist due to the fact that lower incomplete gamma function cannot be approximated if the interference power is too large.

Fig.~\ref{Outage_individual with m=2} shows the outage probability achieved by NOMA users in LoS scenario. In order to better illustrate the performance affected by LoS transmission, the NLoS case is also shown in the figure as a benchmark for comparison. In Fig.~\ref{Outage_individual with m=2}, we can see that as the fading parameter of small scale fading channels, $m$, increases, the outage probability decreases for difference interference power levels. %copy from walidD2Dfig3
This is because that the LoS link between the UAV and users provides higher received power level. It is also shown that the diversity order for the LoS links is one in the case of fixed interference power. This is due to the fact that NOMA users share the same small fading coefficient matrix generated by signal alignment technique. The light blue curves are the outage probability of near users with interference in LoS and in NLoS scenario, respectively. Besides, several observations are drawn as follows: 1) When the interference power level is fixed, the diversity order of the transmission is one in both LoS or NLoS scenarios. 2) For the case that interference power is proportional to the UAV power, diversity orders are zero in both LoS and NLoS scenarios for NOMA users.

Fig. \ref{m2_3D_outage} plots the outage probability of NOMA users in LoS scenario versus the height of the UAV and its target rate. In this case, the transmit SNR is fixed to 60dB. One can observe that an outage ceiling exists even if the height goes to zero. This is because the fact that the target rate of far users is higher than the threshold, which is ${\log _2}\left( {1 + \frac{{\alpha _{k'}^2}}{{\alpha _k^2}}} \right)$.
It is also shown that when the disc radius is fixed, as the distance between the UAV and users increases, the outage probability of near users increases much faster than far users. This is a byproduct of the fact that when the height of the UAV is much larger than disc radius, the large scale fading between the UAV and NOMA users can be considered as the same.
Hence, far users yield better outage probabilities because of higher received power from the UAV.

Fig.~\ref{m2_3D_outage_height_radius} plots the outage probability of paired NOMA users versus both the height of the UAV and the disc radius, with fixed target rate $R_k=R_{k'}=1$BPCU and fixed interference power $P_I=30$dBm. Fig.~\ref{m2_3D_outage_height_radius} clearly illustrates the impact of disc radius to the proposed framework. We can see that the outage probability of near users increases as disc radius increases. It is also worth to mention that when the height goes to zero, the proposed UAV framework is degenerated to the traditional BS framework. Thus, the outage of far users occurs more frequently than near users, which is similar to previous literature. Besides, it is also demonstrated that the outage probabilities of NOMA users are decreased dramatically when decreasing target rates of paired users. The simulation results also confirms the insights from Fig. \ref{m2_3D_outage} that the outage probability of near users is higher than far users in the case of remote UAV.

\begin{figure*}[t!]
\centering
\subfigure[Ergodic rate of NOMA versus transmit SNR in NLoS scenario.]{\label{Fig:Ergodic rate_individual with m=1}
\includegraphics[width =3.1in]{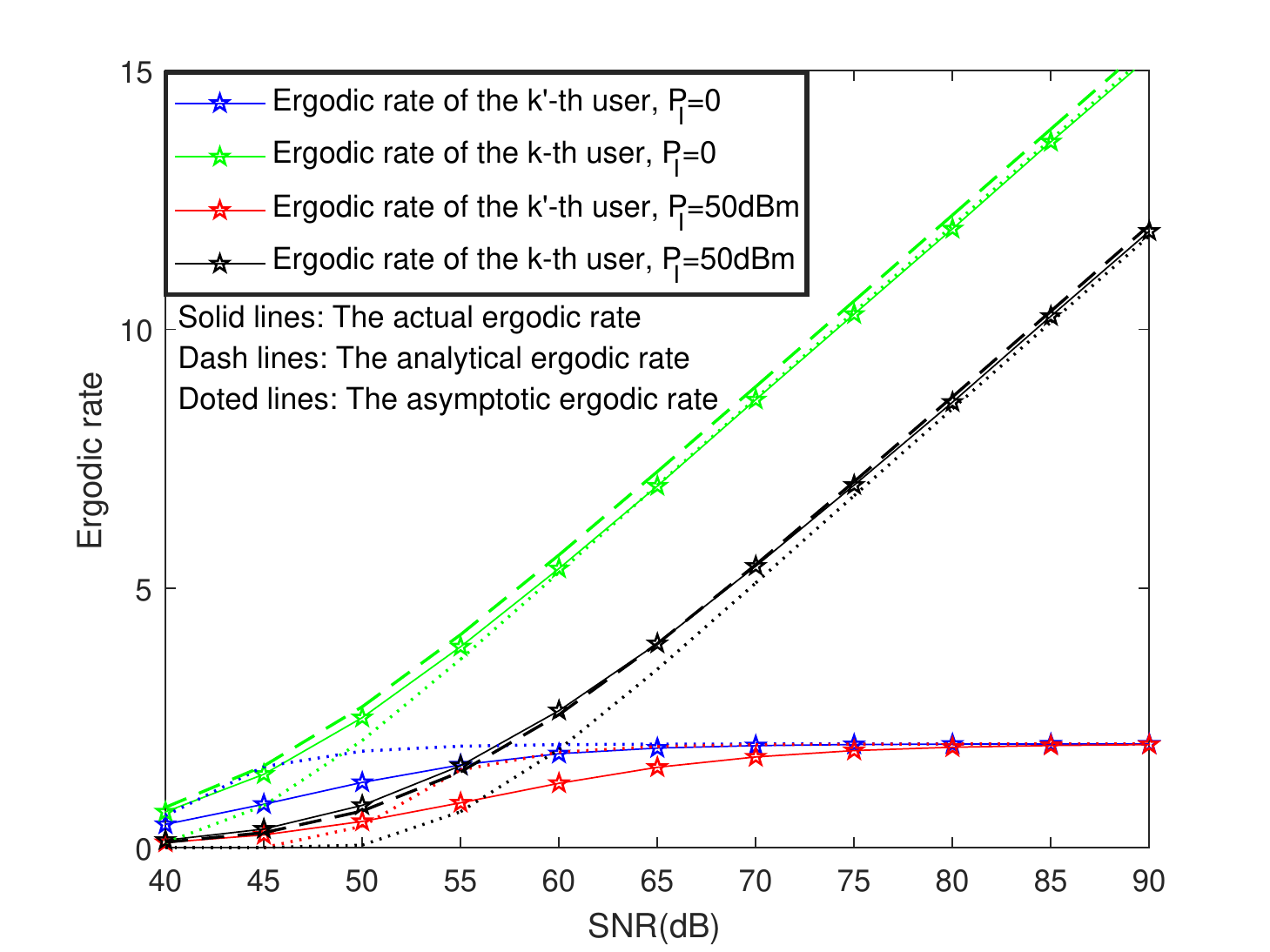}}
\subfigure[Ergodic rate of NOMA versus transmit SNR in LoS scenario.]{\label{Fig:Ergodic rate_individual with m=2}
\includegraphics[width= 3.1in]{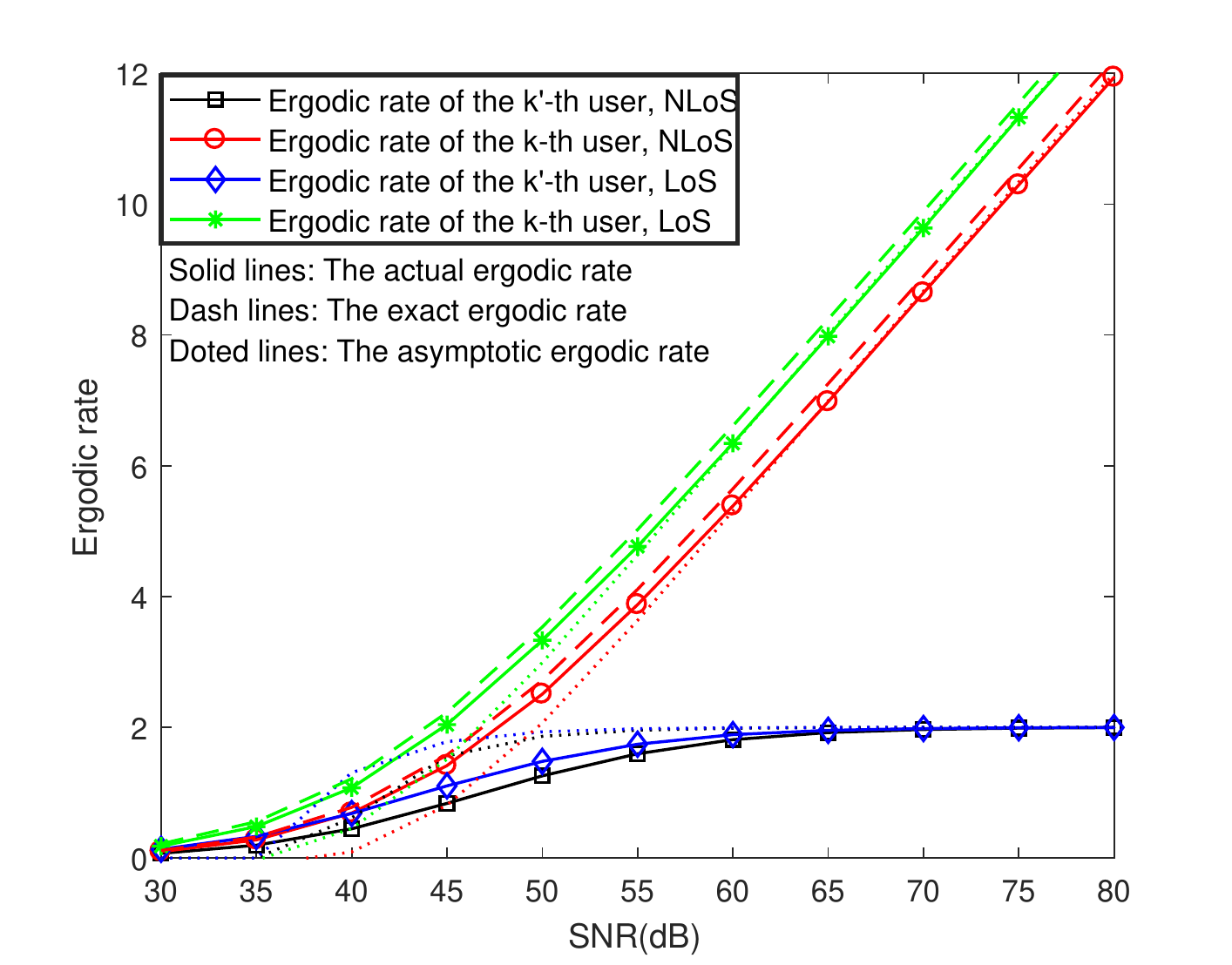}}
\label{Fig4:Ergodic_rate}
\caption{Ergodic rate of NOMA versus transmit SNR. The exact results are calculated from \eqref{exact ergodic m}. The asymptotic results are derived from \eqref{ergodic asymp m' express in paper} and \eqref{m ergodic asym}.}
\vspace{-0.25in}
\end{figure*}
\vspace{-0.1in}
\subsection{Ergodic rates}
\vspace{-0.1in}
Fig. \ref{Fig:Ergodic rate_individual with m=1} compares the ergodic rates of individual users versus transmit SNR with different power levels of interference sources. Several observations can be drawn as follows: 1) An ergodic rate ceiling for far users exists even if the transmit SNR goes to infinity. This is due to the fact that the ergodic rate of far users is entirely affected by power allocation factors of paired NOMA users, which is a constant in F-NOMA. 2) The black line and the red line are the ergodic rate of near users and far users with interference sources, respectively. As transmit SNR increases, the ergodic rate of near users increases because the received signal is increased, which improves the performance of ergodic rate. 3) The asymptotic simulations and analytical simulations are provided to confirm the close agreement between analytical results and the simulations. The dotted lines and dashed lines show the precise agreement between the approximation results and exact results. Thus, for the case of $\alpha=3$, the approximation results and exact results can be considered as the same. 4) As transmit SNR increases, the ergodic rates with and without interference sources for far users are approaching. This is due to the fact that the fixed interference power level is getting relatively smaller when SNR increased, which leads the received signal power relatively higher.

Fig. \ref{Fig:Ergodic rate_individual with m=2} depicts the ergodic rates of individual users versus transmit SNR in LoS scenario. The performance of NLoS scenario is also shown in the figure as a benchmark for comparison. As can be seen from the figure, far users in both NLoS and LoS scenarios have the same ergodic rate ceiling. This phenomenon validates the insight from \textbf{Remark~\ref{remark:ergodic m'}}, where the ergodic rate of far users is only dependent on power allocation coefficients. Besides, the ergodic rate of near users with LoS link is higher than the NLoS situation. This is because the LoS propagation increases the received power level, which increases the ergodic rate of near users.

\begin{figure*}[t!]
\centering
\includegraphics[width= 3.5in]{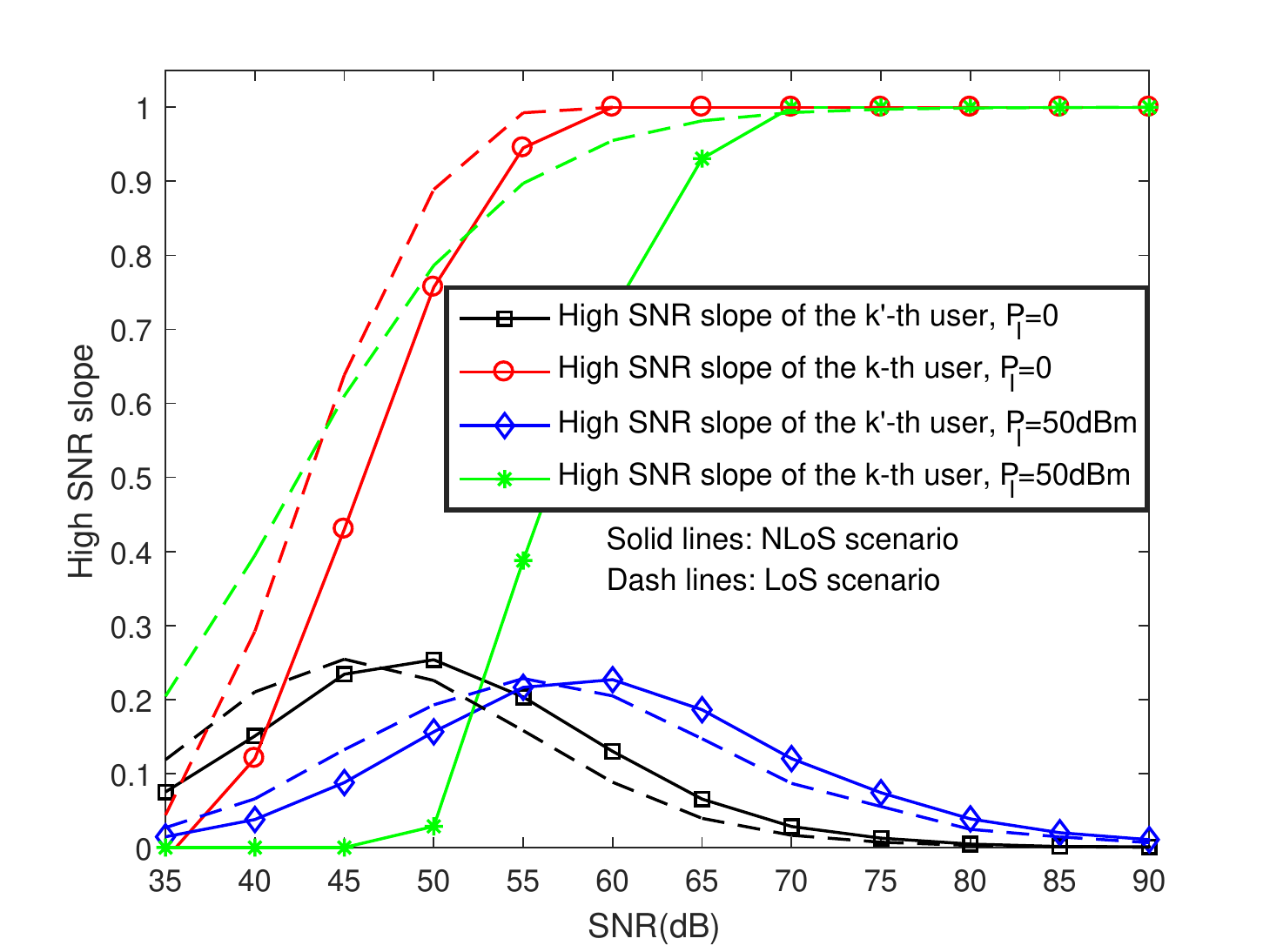}
\caption{High SNR slope of NOMA versus transmit SNR in both LoS and NLoS scenarios.}
\label{Fig:High SNR slope}
\vspace{-0.25in}
\end{figure*}

Fig. \ref{Fig:High SNR slope} plots the high SNR slope of paired NOMA users versus transmit SNR. It is observed that the high SNR slope of far users and near users goes to zero and one, respectively. This behavior can be explained as follows. The ergodic rate of far users, which changes slightly in the high SNR regime, is entirely determined by the power allocation factors. Thus, in the low SNR regime, as $\rho$ increases, the high SNR slope of far users is increased. On the other hand, the high SNR slope is zero for far users in the high SNR regime. For near users, the SNR slope increases monotonously, which shows the high SNR slope for near users is one in the high SNR regime. Another insight is that the LoS link accelerates the increasing rate and the decreasing rate of the paired NOMA users. On the other hand, the slopes of two scenarios are the same, which confirms \textbf{Proposition~\ref{proposition2: high SNR slope}}.
\vspace{-0.15in}
\section{Conclusions}
In this paper, the application of MIMO-NOMA assisted UAV networks was designed. Specifically, stochastic geometry based techniques were used for modeling both the locations of NOMA users and of the interference sources. Additionally, new analytical expressions for outage probability and ergodic rate were derived for characterizing the performance in MIMO-NOMA enhanced UAV frameworks. Diversity orders and high SNR slopes were also obtained to evaluate the system performance. It was analytically demonstrated that the diversity order was one due to the property of signal alignment technique in both LoS and NLoS scenarios. It was also shown that the ergodic rate of far users was only dependent on power allocation factors.

\numberwithin{equation}{section}
\section*{Appendix~A: Proof of Lemma~\ref{lemma1:new channel state}} \label{Appendix:A}
\renewcommand{\theequation}{A.\arabic{equation}}
\setcounter{equation}{0}

In order to estimate the outage probability and ergodic rate of NOMA users, the channel factor ${({\rm \bf B}^{-1}{\rm \bf B}^{-H})}_{k,k}$ can be rewritten as ${({{\rm \bf B}^{ - 1}}{{\rm \bf B}^{ - H}})_{k,k}}{\rm{ = (\rm \bf b}}_k^H({{\rm \bf I}_K} - {\Theta _k}){{\rm \bf b}_k}{)^{ - 1}}$, where ${\Theta _k} = {{\rm \bf \tilde B}_k}{({\rm \bf \tilde B}_k^H{{\rm \bf \tilde  B}_k})^{ - 1}}{\rm \bf \tilde B}_k^H$. Recall that the effective vector ${\rm \bf b}_i$ is generated by (\ref{detection vector design}), i.e.,
\begin{equation}\label{gm}
{{\rm \bf b}_k} = \frac{1}{2}\left[ {\begin{array}{*{20}{c}}
{{\rm \bf {H}}_k^H}&{{\rm \bf {H}}_{k'}^H}
\end{array}} \right]{\left[ {\begin{array}{*{20}{c}}
{{\rm \bf {t}}_k^H}&{{\rm \bf {t}}_{k'}^H}
\end{array}} \right]^H}
= \frac{1}{2}\left[ {\begin{array}{*{20}{c}}
{{\rm \bf {H}}_k^H}&{{\rm \bf {H}}_{k'}^H}
\end{array}} \right]{{\rm \bf {D}}_k}{{\rm \bf{x}}_k},
\end{equation}
where all elements of $\left[ {\begin{array}{*{20}{c}}
{{\rm \bf {H}}_k^H}&{{\rm \bf {H}}_{k'}^H}
\end{array}} \right]$ are independent and identically complex Gaussian distributed, and ${{\rm \bf x}_k}$ is a random vector. Based on two random matrices, we can define two matrix ${{\rm \bf {A}}_k} = \left[ {\begin{array}{*{20}{c}}
{{{\rm \bf {H}}_k}}&{{- }{{\rm \bf {H}}_{k'}}}
\end{array}} \right]$ and ${{\rm \bf {B}}_k} = \left[ {\begin{array}{*{20}{c}}
{{{\rm \bf {H}}_k}}&{{{\rm \bf{H}}_{k'}}}
\end{array}} \right]$. It is quite challenging to estimate the general case. Thus, for the special case that $K=N=1$, and the real-value channel coefficients, two matrices can be written as
\begin{equation}\label{Am}
{{\rm \bf {A}}_k} = \left[ {\begin{array}{*{20}{c}}
{{ h}_k}&{{ { - }}{ h}_{k'}}
\end{array}} \right],
\end{equation}
and
\begin{equation}\label{Bm}
{{\rm \bf {B}}_k} = \left[ {\begin{array}{*{20}{c}}
{{ h}_k}&{{ h}_{k'}}
\end{array}} \right],
\end{equation}
where ${{ h}_k}$ and ${{ h}_{k'}}$ denote the $k$-th and $k'$-th channel coefficient between the UAV and users in the case of $K=N=1$.
Therefore, the base vector ${\rm \bf x}_k$ can be expressed as
\begin{equation}\label{xm}
{{\rm \bf {x}}_k} = \frac{{\left[ {\begin{array}{*{20}{c}}
{{h_{k'}}}&{{h_k}}
\end{array}} \right]}}{{\sqrt {h_k^2 + h_{k'}^2} }}.
\end{equation}
Thus, the variable ${{\rm \bf{D}}_k}{{\rm \bf{x}}_k}$ can be transformed into
\begin{equation}\label{UmXm}
{{\rm \bf {D}}_k}{{\rm \bf {x}}_k} = \frac{{2{h_k}{h_{k'}}}}{{\sqrt {h_k^2 + h_{k'}^2} }}.
\end{equation}

Accordingly, we can rewrite the above matrix by Box-Muller transform \cite{Ding_max_muller,Max_muller_transform} to derive two variable as
\begin{equation}\label{box_muller_trans1}
{h_k} = \sqrt { - 2\ln {W_1}} \cos (2\pi {W_2}),
\end{equation}
and
\begin{equation}\label{box_muller_trans2}
{h_{k'}} = \sqrt { - 2\ln {W_1}} \sin (2\pi {W_2}),
\end{equation}
where $W_1$ and $W_2$ are independent random variables, and uniformly distributed in the interval $\left( {0,m} \right]$. Substituting \eqref{box_muller_trans1} and \eqref{box_muller_trans2} into \eqref{UmXm}, one can obtain
\begin{equation}\label{after box muller}
{{\rm \bf {D}}_k}{{\rm \bf{x}}_k} = \sqrt { - 2\ln {W_1}} \sin (2\pi {W_2})\cos (2\pi {W_2})=\sqrt { - 2\ln {W_1}} \sin (4\pi {W_2}).
\end{equation}
Replacing $\sin (4\pi {W_2})$ with $\sin (2\pi {W_2})$ will not change the density function because $W_2$ is also uniformly distributed in the interval $\left( {0,m} \right]$. Therefore, we can consider the following distribution holds
\begin{equation}\label{conclusion}
{{\rm \bf {D}}_k}{{\rm \bf {x}}_k}  \sim {\rm {\mathcal{CN}}} \left( {0,m {\rm \bf I}_K} \right).
\end{equation}

Hence, ${\rm \bf H}_k$ is a $K \times 1$ complex Gaussian vector with zero mean and $m$ variance, and the lemma is proved.

\numberwithin{equation}{section}
\section*{Appendix~B: Proof of Theorem~\ref{Theorem1:Outage m' upper}} \label{Appendix:Bs}
\renewcommand{\theequation}{B.\arabic{equation}}
\setcounter{equation}{0}

First, the upper bound of outage probability can be written as follows:
\begin{equation}\label{new outage}
{\rm \hat P}_{{k'}}^O = \left( {\frac{{\rho {{\left| {{u_{k'}}} \right|}^2}\alpha _{{k'}}^2}}{{2 + \rho {{\left| {{u_{k'}}} \right|}^2}\alpha _k^2 + 2\delta \sum\limits_{{j} \in {\Psi _I}} {{P_I}d_{j,k'}^{ - \alpha }} }} < {\varepsilon _{k'}}} \right),
\end{equation}
where ${\varepsilon _{k'}} = {2^{{R_{k'}}}} - 1$. After some algebraic manipulations, the above function can be transformed into
\begin{equation}\label{transformed outage}
{\rm{\hat P}}_{{k'}}^O = \left( {2\frac{{{\varepsilon _{k'}}}}{{  {\rho \alpha _{{k'}}^2 - \rho \alpha _{k}^2{\varepsilon _{k'}}} }} + 2\delta {P_I}\sum\limits_{{j} \in {\Psi _I}} {d_{j,k'}^{ - \alpha }} \frac{{{\varepsilon _{k'}}}}{{ {\rho \alpha _{{k'}}^2 - \rho \alpha _{k}^2{\varepsilon _{k'}}} }} > {{\left| {{u_{k'}}} \right|}^2}} \right),
\end{equation}
if $\rho \alpha _{{k'}}^2 - \rho \alpha _{k}^2{\varepsilon _{k'}} > 0$. Denote ${V_{k'}} = \frac{{{\varepsilon _{k'}}}}{{  {\rho \alpha _{{k'}}^2 - \rho \alpha _{k}^2{\varepsilon _{k'}}} }}$, which is a constant in F-NOMA.
Recall that $\frac{1}{{{{({{\rm \bf B}^{ - 1}}{{\rm \bf B}^{ - H}})}_{k,k}}}}$ is exponentially distributed. Therefore, the upper bound of outage probability can be expressed as
\begin{equation}\label{outage prob of gauss}
{\rm \hat P}_{{k'}}^O = {\rm{1}} - {{\rm \mathbb E}_{d_{j,k'},{d_{k'}}}}\left\{ {{\rm{exp}}\left( {{ - {\frac{2}{m}}{V_{k'}}d_{k'}^\alpha } { - {\frac{2}{m}}{V_{k'}}\delta {P_I}d_{k'}^\alpha \sum\limits_{{j} \in {\Psi _I}} {d_{j,k'}^{ - \alpha }} } } \right)} \right\},
\end{equation}
where $\rho \alpha _{{k'}}^2 - \rho \alpha _{k}^2{\varepsilon _{k'}} > 0$.

In order to deduct the outage probability, we first focus on the interference sources, which are allocated under HPPP distribution with density $\lambda_I$. The HPPP distribution $\Psi _I$ can be considered as stationary, and hence the summation of interference sources can be equivalently calculated by focusing on the interference reception seen at the node located at the $k'$-th user. Accordingly, the expectation of interference sources can be obtain as by Laplace functionals \cite{Stochastic_Geo}:
\begin{equation}\label{interferences}
\begin{aligned}
{\rm \mathbb E} \left\{ {{d_{j,k'}}} \right\} &= {\rm \mathbb E}\left\{ {{\rm{exp}}\left( { - {\frac{2}{m}}{V_{k'}}\delta {P_I}d_{k'}^\alpha \sum\limits_{{j} \in {\Psi _I}} {d_{j,k'}^{ - \alpha }} } \right)} \right\} \\
&= {\rm{exp}}\left( { - {\lambda _I}\int\limits_{{j} \in \mathbb{R} } { {1 - \exp \left( { - {\frac{2}{m}}{V_{k'}}\delta {P_I}d_{k'}^\alpha j^\alpha } \right)dj}} } \right),
\end{aligned}
\end{equation}
where $j$ denotes the interferers allocated in the disc $\mathbb{R}$. For notation simplicity, ${\phi _{k'}}$ denotes $\frac{2}{m}{V_{k'}}\delta {P_I}d_{k'}^\alpha$. It is assumed that the interference sources does not distributed in the disc $r_0$ to prevent infinite received signal power. After changing to polar coordinates, the above expectation can be converted as
\begin{equation}\label{Expectation of I}
Q_{j,k'}={\rm \mathbb E} \left\{ {{d_{j,k'}}} \right\}  = \exp \left( { - {\lambda _I}\pi \phi _{k'}^{\frac{2}{\alpha }}\gamma \left( {\begin{array}{*{20}{c}}
{\frac{1}{\alpha },}&{\frac{{{\phi _{k'}}}}{{r_0^\alpha }}}
\end{array}} \right)} \right).
\end{equation}

Then the outage probability can be rewritten as
\begin{equation}\label{outage distance from m'}
{\rm{\hat P}}_{{k'}}^O = {\rm{1}} - {{\rm \mathbb E}_{{d_{k'}}}}\left\{ {{\rm{exp}} {\left( { - {\frac{2}{m}}{V_{k'}}d_{k'}^\alpha } \right)Q_{j,k'}} } \right\}.
\end{equation}

Recall that in this article, the $k'$-th user, which has poorer channel gain, is uniformly distributed in the coverage disc. Therefore, we can generally consider that the horizontal distance of user $k'$ is greater than user $k$, and the outage probability can be transformed into
\begin{equation}\label{outage before result}
{\rm{\hat P}}_{{k'}}^O = {\rm{1}} - \frac{1}{{\pi R_d^2 - \pi R_m^2}}\int\limits_{p \in D} { {\exp \left( { - {\frac{2}{m}}{V_{k'}}d_{k'}^\alpha } \right)Q_{j,k'}dp} },
\end{equation}
where $p$ denotes the users location of the $k'$-th user. Again, the polar coordination for above equation is shown as
\begin{equation}\label{origin expression on ground}
{\rm{\hat P}}_{k'}^O = {\rm{1}} - \frac{2}{{R_d^2 - R_m^2}}\int_{{R_m}}^{{R_d}} { {\exp \left( { - \frac{2}{w}{V_{k'}}{{{\left( {{r^2} + {h^2}} \right)}^{\frac{\alpha }{2}}}}} \right){Q_{j,k'}}rdr}}.
\end{equation}

Using Riemann--Stieltjes integral, the above integral can be further transformed into

\begin{equation}\label{outage uper bounds}
{\rm{\hat P}}_{{k'}}^O = {\rm{1}} - \frac{2}{{R_d^2 - R_m^2}}\int_{{l_{{m}}} }^{{l_{{d}}} } { {\exp \left( { - {\frac{2}{m}}{V_{k'}}{{x}}_{}^\alpha } \right)Q_{j,k'}xdx} },
\end{equation}
where ${l_{{d}}} = \sqrt {R_d^2 + {h^2}}$, and ${l_{{m}}} = \sqrt {R_m^2 + {h^2}}$. Hence, the theorem is proved.

\numberwithin{equation}{section}
\section*{Appendix~C: Proof of Corollary~\ref{Corollary1: m' asymptotic}} \label{Appendix:Cs}
\renewcommand{\theequation}{C.\arabic{equation}}
\setcounter{equation}{0}

The asymptotic result of user $k'$ is worth estimating. In the asymptotic outage probability, the transmit SNR between the UAV and users obeys ${\rho  \to \infty }$. In this case, the lower incomplete Gamma function can be evaluated using the power series expansion to
\begin{equation}\label{Gamma expand}
\gamma \left( {\begin{array}{*{20}{c}}
{\frac{1}{\alpha },}&{\frac{{{\phi _{k'}}}}{{r_0^\alpha }}}
\end{array}} \right) = \sum\limits_{n = 0}^\infty  {\frac{{{{\left( { - 1} \right)}^n}{{\left( {\frac{{{\phi _{k'}}}}{{r_0^\alpha }}} \right)}^{\left( {\frac{1}{\alpha } + n} \right)}}}}{{n!\left( {\frac{1}{\alpha } + n} \right)}}}  \approx \alpha {\left( {\frac{{{\phi _{k'}}}}{{r_0^\alpha }}} \right)^{\frac{1}{\alpha }}}.
\end{equation}

Note that $\mathop {\lim }\limits_{x \to 0 } \left( {1 - {e^{ - x}}} \right) \approx x$, and the upper bound of the outage probability can be approximated at high transmit SNR regime as follows:
\begin{equation}\label{asymptotic analytical result}
\begin{aligned}
{\rm{\hat P}}_{{k'}}^\infty &= {\rm{1}} - \frac{2}{{R_d^2 - R_m^2}}\int_{\sqrt {{h ^2} + R_m^2} }^{\sqrt {{h ^2} + R_d^2} } {\left( {\exp \left( { - {\frac{2}{m}}{V_{k'}}x_{}^\alpha } \right)\exp \left( { - {\frac{2}{m}}{V_{k'}}\pi {\lambda _I}\delta {P _I}\frac{\alpha }{{{r_0}}}x_{}^\alpha } \right)xdx} \right)}\\
& = 1 - \frac{2}{{R_d^2 - R_m^2}}\int_{\sqrt {{h ^2} + R_m^2} }^{\sqrt {{h ^2} + R_d^2} } {\left( {x - {T_{k'}}{x^{\alpha  + 1}}} \right)} dx
 = \frac{{2{T_{k'}}}}{{R_d^2 - R_m^2}} \frac{{{{({h ^2} + R_d^2)}^{\frac{{\alpha  + 2}}{2}}} - {{({h ^2} + R_m^2)}^{\frac{{\alpha  + 2}}{2}}}}}{{\alpha  + 2}} ,
\end{aligned}
\end{equation}
where ${T_{k'}} = {\frac{2}{m}}{V_{k'}}\left( {1 + \pi {\lambda _I}\delta {P_I}\frac{\alpha }{{{r_0}}}} \right)$, and the corollary is proved.

\numberwithin{equation}{section}
\section*{Appendix~D: Proof of Theorem~\ref{Theorem2:Outage m upper}} \label{Appendix:Ds}
\renewcommand{\theequation}{D.\arabic{equation}}
\setcounter{equation}{0}

First, we rewrite the upper bound of outage probability as follows:
\begin{equation}\label{upper bound m}
{\rm{\hat P}}_k^O = {\rm{P}}\left( {{{\left| {{u_k}} \right|}^2} < {\frac{2}{m}}{V_{k'}}{\varphi _k}} \right) + {\rm{P}}\left( {{\frac{2}{m}}{V_{k'}}{\varphi _k} < {{\left| {{u_k}} \right|}^2} < {\frac{2}{m}}{V_k}{\varphi _k}} \right),
\end{equation}
if $\rho \alpha _{{k'}}^2 - \rho \alpha _{k}^2{\varepsilon _{k'}} > 0$. In the downlink scenario, it is assumed that the $k$-th user is the user with higher channel gain. Thus, the above outage probability can be further transformed into
\begin{equation}\label{more restriction}
{\rm{\hat P}}_k^O = {\rm{P}}\left( {{{\left| {{u_k}} \right|}^2} < {\frac{2}{m}}{\varphi _k} \cdot \max \left\{ {{V_{k'}},{V_k}} \right\}} \right).
\end{equation}

The interference sources are under the same HPPP distribution, and user $k$ is uniformly distributed in the small disc. Therefore, similar to the arguments from~\eqref{outage prob of gauss} to~\eqref{outage uper bounds}, the outage probability can be calculated as follows:
\begin{equation}\label{outage m}
{\rm{\hat P}}_{k}^O = {\rm{1}} - \frac{2}{{R_m^2}}\int_h^{{l_{{m}}}} { {\exp \left( { - {\frac{2}{m}}\max \left\{ {{V_{k'}},{V_k}} \right\}{x^\alpha }} \right)Q_{j,k}xdx} } .
\end{equation}
Hence, the theorem is proved.

\numberwithin{equation}{section}
\section*{Appendix~E: Proof of Corollary~\ref{Corollary3: m asymptotic}} \label{Appendix:Es}
\renewcommand{\theequation}{E.\arabic{equation}}
\setcounter{equation}{0}

To obtain the ergodic rate of the far user, the expectation of the ergodic rate is expressed as
\begin{equation}\label{ergodic expression}
{R_{k'}} = \mathbb{E}\left\{ {{{\log }_2}\left( {1 + SIN{R_{k'}}\left( {x_{k'}} \right)} \right)} \right\}.
\end{equation}
After some algebraic manipulations, we have
\begin{equation}\label{ergodic after}
\begin{aligned}
{\mathbb{E} \left\{ {{{\log }_2}\left( {1 + SIN{R_{k'}}\left( {x_{k'}} \right)} \right)} \right\}} &=  - \int\limits_0^\infty  {{{\log }_2}(1 + {x_{k'}})} d\left( {1 - F\left( {{x_{k'}}} \right)} \right)\\
& = \frac{1}{{\ln \left( 2 \right)}}\int\limits_0^\infty  {\frac{{1 - F\left( {{x_{k'}}} \right)}}{{1 + {x_{k'}}}}} d{x_{k'}}.
\end{aligned}
\end{equation}

In this paper, NOMA users are modeled according to HPPPs. Therefore, the users are independently and identically distributed in the coverage area, and the probability density functions (PDFs) of two NOMA users are given by
%${F\left( x \right)}$ is an exponential distribution. Thus, $F\left( x \right) = 1 - \exp ( - {\vartheta _{m'}}x)$,
\begin{equation}\label{PDF of m'}
{f_{k'}}\left( x \right) = \frac{{{\lambda_{k'}}{\Psi_{k'}}}}{{{\lambda _{k'}}{{\Psi}_{k'}}\left( {\pi R_d^2 - \pi R_m^2} \right)}} = \frac{1}{{\pi R_d^2 - \pi R_m^2}},
\end{equation}
and
\begin{equation}\label{PDF of m}
{f_k}\left( x \right) = \frac{1}{{\pi R_m^2}},
\end{equation}
respectively.
Therefore, the cumulative distribution function (CDF) of the far user $k'$ can be calculated as
\begin{equation}\label{CDF of m'}
{F_{{x_{k'}}}}\left( {{x_{k'}}} \right) = \frac{2}{{R_d^2 - R_m^2}}\int_{{l_m}}^{{l_d}} { { \exp \left( { - {\frac{2}{m}}{\varepsilon _{{x_{k'}}}}\left( {1 + \delta {{\rm \bf I}_{{{k'}}}} } \right)}{r^\alpha } \right)} } rdr,
\end{equation}
if ${x_{k'}} < \frac{{\alpha _{k'}^2}}{{\alpha _k^2}} \buildrel \Delta \over = \hat \alpha $, and ${\varepsilon _{{x_{k'}}}}$ denotes $\frac{{{x_{k'}}}}{{\rho \alpha _{{k'}}^2 - \rho \alpha _{k}^2{x_{k'}}}}$.

Substituting~\eqref{CDF of m'} into~\eqref{ergodic after}, the expectation rate of the $k'$-th user can be rewritten as
\begin{equation}\label{results}
\begin{aligned}
& {\mathbb{E} \left\{ {{{\log }_2} \left( {1 + SIN{R_{k'}}\left( x \right)} \right)} \right\}}  \\
&= \frac{1}{{\ln \left( 2 \right)}}\int\limits_0^{\hat \alpha}  {\frac{{1 - \frac{2}{{R_d^2 - R_m^2}}\int_{{l_m}}^{{l_d}} {{ \exp \left( { - {\frac{2}{m}}{\varepsilon _{{x_{k'}}}}\left( {1 + \delta {{\rm \bf I}_{{{k'}}}} } \right)}{r^\alpha } \right)} } rdr}}{{1 + {x_{k'}}}}} d{x_{k'}}.
\end{aligned}
\end{equation}

One can know from~\eqref{non interference existed} that the exact expression of CDF includes lower incomplete Gamma function, which is challenging to calculate. Therefore, in this paper, we only estimate the approximated ergodic rate, and the ergodic rate of user $k'$ can be expressed as
\begin{equation}\label{asym erogic rate}
{R_{k'}} = \frac{1}{{\ln \left( 2 \right)}}\int\limits_0^{\hat \alpha } {\frac{1}{{1 + {x_{k'}}}}} d{x_{k'}} + \frac{1}{{\ln \left( 2 \right)}}\int\limits_0^{\hat \alpha } {\frac{{{Q_{k'}}{x_{k'}}}}{{\left( {1 + {x_{k'}}} \right)\left( {\alpha _{{k'}}^2 - \alpha _{k}^2{x_{k'}}} \right)}}} d{x_{k'}},
\end{equation}
where ${Q_{k'}} = \frac{{4\left( {1 + \frac{\pi {\lambda _I}\delta {P_I} }{{{\alpha r_0}}}}\right)\left( {l_d^{\alpha  + 2} - l_m^{\alpha  + 2}} \right)}}{{w\rho \left( {R_d^2 - R_m^2} \right)\left( {\alpha  + 2} \right)}}$.

After some algebraic manipulations,~\eqref{asym erogic rate} can be obtained as follows:
\begin{equation}\label{ergodic asymp express}
{R_{k'}} = {\log _2}(1 + \hat \alpha ) + {Q_{k'}}\hat \alpha {\log _2}(1 + {\alpha _{k}^2}) - {Q_{k'}}{\log _2}(1 + \hat \alpha ).
\end{equation}
Thus, the proof of corollary is completed.
\vspace{-0.1in}
\numberwithin{equation}{section}
\section*{Appendix~F: Proof of Theorem~\ref{Theorem4:m ergodic rate}} \label{Appendix:Fs}
\renewcommand{\theequation}{F.\arabic{equation}}
\setcounter{equation}{0}

The proof start by providing the ergodic rate of the near user $k$ as follows:
\begin{equation}\label{m ergodic exp}
{R_k} = {\mathbb E}\left\{ {{{\log }_2}\left( {1 + SIN{R_k}\left( {{x_k}} \right)} \right)} \right\}
\end{equation}

Similar to the steps from~\eqref{ergodic after} to~\eqref{asym erogic rate}, the ergodic rate can be illustrated as
\begin{equation}\label{expression}
{R_k} = \frac{1}{{\ln \left( 2 \right)}}\int\limits_0^\infty  {\frac{{\frac{2}{{R_m^2}}\int_h^{{l_m}} {\left( {\exp \left( { - \frac{2}{m}{\varepsilon _{{x_k}}}\left( {1 + \delta {{\rm \bf I}_{{{k}}}}} \right){r^\alpha }} \right)} \right)} rdr}}{{1 + {x_k}}}} d{x_k},
\end{equation}
where ${\varepsilon _{{x_k}}} = \frac{{{x_k}}}{{\rho \alpha _{k}^2}}$.
After some algebraic manipulations, the ergodic rate can be transformed into

\begin{equation}\label{full express}
\begin{aligned}
{R_k} &= \frac{{l_m^2}}{{\ln \left( 2 \right)R_m^2}}\underbrace {\int\limits_0^\infty  {\frac{{{e^{{\rm{ - }}Cl_m^\alpha {x_k}}}}}{{1 + {x_k}}}} d{x_k}}_{{J_1}} - \frac{{{h^2}}}{{\ln \left( 2 \right)R_m^2}}\underbrace {\int\limits_0^\infty  {\frac{{{e^{{\rm{ - }}C{h^\alpha}{x_k}}}}}{{1 + {x_k}}}} d{x_k}}_{{J_2}} \\
&- \frac{{{C^{ - \frac{2}{\alpha }}}}}{{\ln \left( 2 \right)R_m^2}}\underbrace {\int\limits_0^\infty  {\frac{{{x_k^{ - \frac{2}{\alpha }}}\gamma \left( {\frac{2}{\alpha } + 1,Cl_m^\alpha{x_k}} \right)}}{{1 + {x_k}}}} d{x_k}}_{{J_3}} + \frac{{{C^{ - \frac{2}{\alpha }}}}}{{\ln \left( 2 \right)R_m^2}}\underbrace {\int\limits_0^\infty  {\frac{{{x_k^{ - \frac{2}{\alpha }}}\gamma \left( {\frac{2}{\alpha } + 1,C{h^\alpha}{x_k}} \right)}}{{1 + {x_k}}}} d{x_k}}_{{J_4}},
\end{aligned}
\end{equation}
where $C = \frac{{2\left( {1 + \frac{\pi {\lambda _I}\delta {P _I}}{\alpha r_0}  } \right)}}{{m\rho \alpha _{k}^2}}$.

Based on \cite[eq. (3.352.4)]{Table_of_integrals} and applying polynomial expansion manipulations, $J_1$ and $J_2$ can be expressed as
\begin{equation}\label{J1}
{J_1} = \frac{{{\rm{ - }}l_m^2}}{{\ln \left( 2 \right)R_m^2}}{e^{{\rm{  }}Cl_m^\alpha }}{\rm{Ei}}\left( -{Cl_m^\alpha } \right),
\end{equation}
and
\begin{equation}\label{J2}
{J_2} = \frac{{{\rm{ - }}{h^2}}}{{\ln \left( 2 \right)R_m^2}}{e^{{\rm{  }}C{h^\alpha }}}{\rm{Ei}}\left( -{C{h^\alpha }} \right).
\end{equation}

Calculating $J_3$ and $J_4$ is challenging due to the lower incomplete Gamma function. Thus, the lower incomplete Gamma function can be evaluated using the exponential series expansion to $\gamma \left( {\frac{2}{\alpha } + 1,Cl_m^\alpha {x_k}} \right) = \sum\limits_{p = 0}^\infty  {\frac{{{{\left( {Cl_m^\alpha {x_m}} \right)}^{\frac{2}{\alpha } + 1 + p}}{e^{ - Cl_m^\alpha {x_k}}}}}{{\left( {\frac{2}{\alpha } + 1} \right) \ldots \left( {\frac{2}{\alpha } + 1 + p} \right)}}}$. Therefore, $J_3$ and $J_4$ can be rewritten as
\begin{equation}\label{J3}
{J_3} = \sum\limits_{p = 0}^\infty  {\frac{{{{\left( {l_m^\alpha } \right)}^{\frac{2}{\alpha } + 1 + p}}{C^{1 + p}}}}{{\left( {\frac{2}{\alpha } + 1} \right) \cdot  \ldots  \cdot \left( {\frac{2}{\alpha } + 1 + p} \right)\ln \left( 2 \right)R_m^2}}} \int\limits_0^\infty  {\frac{{x_k^{1 + p}{e^{ - Cl_m^\alpha {x_k}}}}}{{1 + {x_k}}}} d{x_k},
\end{equation}
and
\begin{equation}\label{J4_final}
{J_4} = \sum\limits_{p = 0}^\infty  {\frac{{{{\left( {{h^\alpha }} \right)}^{\frac{2}{\alpha } + 1 + p}}{C^{1 + p}}}}{{\left( {\frac{2}{\alpha } + 1} \right) \cdot  \ldots  \cdot \left( {\frac{2}{\alpha } + 1 + p} \right)\ln \left( 2 \right)R_m^2}}} \int\limits_0^\infty  {\frac{{x_k^{1 + p}{e^{ - C{h^\alpha }{x_k}}}}}{{1 + {x_k}}}} d{x_k}.
\end{equation}
After some algebraic manipulations, the above integrals can be further transformed into
\begin{equation}\label{J3_final}
\begin{aligned}
{J_3} =& \underbrace { \sum\limits_{p = 0}^\infty  {\frac{{{{\left( {l_m^\alpha } \right)}^{\frac{2}{\alpha } + 1 + p}}{C^{1 + p}}}}{{\left( {\frac{2}{\alpha } + 1} \right) \cdot  \ldots  \cdot \left( {\frac{2}{\alpha } + 1 + p} \right)\ln \left( 2 \right)R_m^2}}} {e^{Cl_m^\alpha }}{\rm{Ei}}\left( { - Cl_m^\alpha } \right)}_{J_{3_l}} \\
&+ \sum\limits_{p = 0}^\infty  {\sum\limits_{i = 1}^{p + 1} {\frac{{{{\alpha \left( { - 1} \right)}^{i - 1}}\left( {i - 1} \right)!l_m^{\alpha p + \alpha  + 2 - \alpha i}{C^{p + 1 - i}}}}{{\left( {\frac{2}{\alpha } + 1} \right) \cdot  \ldots  \cdot \left( {\frac{2}{\alpha } + 1 + p} \right)\ln \left( 2 \right)R_m^2}}} } ,
\end{aligned}
\end{equation}
and
\begin{equation}\label{J4_final}
\begin{aligned}
{J_4} =& \underbrace { \sum\limits_{p = 0}^\infty  {\frac{{{{\left( {{h^\alpha }} \right)}^{\frac{2}{\alpha } + 1 + p}}{C^{1 + p}}}}{{\left( {\frac{2}{\alpha } + 1} \right) \cdot  \ldots  \cdot \left( {\frac{2}{\alpha } + 1 + p} \right)\ln \left( 2 \right)R_m^2}}} {e^{Cl_m^\alpha }}{\rm{Ei}}\left( { - C{h^\alpha }} \right)}_{J_{4_l}} \\
&+ \sum\limits_{p = 0}^\infty  {\sum\limits_{i = 1}^{p + 1} {\frac{{{{\alpha \left( { - 1} \right)}^{i - 1}}\left( {i - 1} \right)!{h^{\alpha p + \alpha  + 2 - \alpha i}}{C^{p + 1 - i}}}}{{\left( {\frac{2}{\alpha } + 1} \right) \cdot  \ldots  \cdot \left( {\frac{2}{\alpha } + 1 + p} \right)\ln \left( 2 \right)R_m^2}}} } .
\end{aligned}
\end{equation}

In the high SNR regime, $J_{3_l}-J_{4_l}$ can be eliminated, and the proof of theorem is completed.

\numberwithin{equation}{section}
\section*{Appendix~G: Proof of Corollary~\ref{Corollary4: m ergodic when a=3}} \label{Appendix:Gs}
\renewcommand{\theequation}{G.\arabic{equation}}
\setcounter{equation}{0}

For the special case of $\alpha=3$, the lower incomplete Gamma function can be further transformed into
\begin{equation}\label{lower gamma to G}
\gamma \left( {\frac{2}{\alpha } + 1,Cl_m^3{x_k}} \right) = G
\begin{tiny}
\begin{array}{*{20}{c}}
{1,1}\\
{1,2}
\end{array}
\end{tiny}
\left( \begin{array}{l}
1\\
\frac{2}{3 } + 1,0
\end{array} \bigg| {Cl_m^3{x_k}}  {} \right),
\end{equation}
where $G\left(  \cdot  \right)$ represents Meijer-G function. Thus, $J_3$ can be deducted to
\begin{equation}\label{J3_Gfunc}
\begin{aligned}
{J_3} &= \frac{{{C^{ - \frac{2}{3}}}}}{{\ln \left( 2 \right)R_m^2}}\int\limits_0^\infty  {\frac{{{x_k^{ - \frac{2}{3}}}G
\begin{tiny}
\begin{array}{*{20}{c}}
{1,1}\\
{1,2}
\end{array}
\end{tiny}
\left( \begin{array}{l}
1\\
\frac{2}{3 } + 1,0
\end{array} \bigg| {Cl_m^3{x_k}} \right)}}{{1 + {x_k}}}} d{x_k}=\frac{{{C^{ -1}l_m^{- 1}}}}{{\ln \left( 2 \right)R_m^2}}G
\begin{tiny}
\begin{array}{*{20}{c}}
{2,2}\\
{3,2}
\end{array}
\end{tiny}
\left( \begin{array}{l}
-1,0,-\frac{2}{3}\\
-\frac{1}{3},0
\end{array} \bigg| \frac{1}{{Cl_m^3}} \right).
\end{aligned}
\end{equation}

Similar to the arguments of \eqref{lower gamma to G} and \eqref{J3_Gfunc}, the exact expression of $J_4$ can be obtained as
\begin{equation}\label{J4_exact}
{J_4} =\frac{{{C^{ -1}h^{- 1}}}}{{\ln \left( 2 \right)R_m^2}}G
\begin{tiny}
\begin{array}{*{20}{c}}
{2,2}\\
{3,2}
\end{array}
\end{tiny}
\left( \begin{array}{l}
-1,0,-\frac{2}{3}\\
-\frac{1}{3},0
\end{array} \bigg| \frac{1}{{Ch^3}} \right).
\end{equation}
Thus, the corollary is proved.

\bibliographystyle{IEEEtran}%
\bibliography{IEEEabrv,bib2018}

%\begin{IEEEbiography}[{\includegraphics[width=1in, height=1.25in, clip, keepaspectratio] {ZhengyuSong.eps}}] {Zhengyu Song}
%received the B.Sc. and M.Sc. degrees from Beijing Jiaotong University and the Ph.D. degree from Beijing Institute of Technology, Beijing, China, all in Information and Communication Engineering. He is currently with the School of Electronic and Information Engineering, Beijing Jiaotong University, Beijing, China. His research interests lie in cooperative communications and radio resource management in multicarrier systems.
%\end{IEEEbiography}

%\begin{IEEEbiography}[{\includegraphics[width=1in, height=1.25in, clip, keepaspectratio] {XinSun.eps}}] {Xin Sun}
%received the Ph.D. degree in electromagnetic measurement technology and instrument from Harbin Institute of Technology, Harbin, China. She is currently a Professor with the School of Electronic and Information Engineering, Beijing Jiaotong University, Beijing, China. Her main research interests are professional mobile communications and wireless personal communications.
%\end{IEEEbiography}

\end{document}